\documentclass[prc,twocolumn,showpacs,superscriptaddress,floatfix]{revtex4}
\usepackage{graphicx}
\usepackage{color}
\usepackage{hyperref}
\usepackage{multirow}
\usepackage{gensymb}
\usepackage{textcomp}
\usepackage{amsfonts}
\usepackage{amsmath}
\usepackage{bm}

\definecolor{grey}{rgb}{.65,.65,.65}
\newcommand{\add}[1]{\color{blue}#1\color{black}{}}

\newcommand{\coloruse}{\marginpar{\scriptsize
\add%
}}

\begin{document}

\title{Electromagnon in Y-type hexaferrite $\textbf{BaSrCoZnFe}_{\textbf{11}}\textbf{Al}\textbf{O}_{\textbf{22}}$}

\author{Jakub V\'\i t}
\thanks{Authors to whom correspondence should be addressed}
\email[e-mail: ]{vit@fzu.cz; kamba@fzu.cz}
\affiliation{Institute of Physics of the Czech Academy of Sciences, Na Slovance 2, 182 21 Prague~8, Czech Republic}
\affiliation{Faculty of Nuclear Science and Physical Engineering, Czech Technical University, B\v{r}ehov\'{a} 7, 115 19 Prague~1, Czech Republic}
\affiliation{Department of Physics, Budapest University of Technology and Economics and MTA-BME Lend\"{u}let Magneto-optical
Spectroscopy Research Group, 1111 Budapest, Hungary}
\author{Filip Kadlec}
\affiliation{Institute of Physics of the Czech Academy of Sciences, Na Slovance 2, 182 21 Prague~8, Czech Republic}
\author{Christelle Kadlec}
\affiliation{Institute of Physics of the Czech Academy of Sciences, Na Slovance 2, 182 21 Prague~8, Czech Republic}
\author{Fedir Borodavka}
\affiliation{Institute of Physics of the Czech Academy of Sciences, Na Slovance 2, 182 21 Prague~8, Czech Republic}
\author{Yi Sheng Chai}
\affiliation{Institute of Physics, Chinese Academy of Sciences, Beijing 100190, P. R. China}
\affiliation{Department of Applied Physics, Chongqing University, Chongqing, 401331, P. R. China}
\author{Kun Zhai}
\affiliation{Institute of Physics, Chinese Academy of Sciences, Beijing 100190, P. R. China}
\author{Young Sun}
\affiliation{Institute of Physics, Chinese Academy of Sciences, Beijing 100190, P. R. China}
\author{Stanislav Kamba}
\affiliation{Institute of Physics of the Czech Academy of Sciences, Na Slovance 2, 182 21 Prague~8, Czech Republic}

\coloruse
\begin{abstract}
We investigated static and dynamic magnetoelectric properties of single crystalline BaSrCoZnFe$_{11}$AlO$_{22}$ which is a room-temperature multiferroic with Y-type hexaferrite crystal structure.  Below $300\,\rm K$, a purely electric-dipole-active electromagnon at $\approx 1.2\,\rm THz$ with the electric polarization oscillating along the hexagonal axis was observed by THz and Raman spectroscopies. We investigated the behavior of the electromagnon with applied DC magnetic field  and linked its properties to static measurements of the magnetic structure. Our analytical calculations determined selection rules for electromagnons activated by the magnetostriction mechanism in various magnetic structures of Y-type hexaferrite. Comparison with our experiment supports that the electromagnon is indeed activated by the magnetostriction mechanism involving spin vibrations along the hexagonal axis.
\end{abstract}
\date{\today}
\pacs{75.85.+t, 75.40.Gb, 78.30.-j}

\maketitle \section{Introduction}

Magnetoelectric (ME) multiferroics are fascinating materials due to a potential
possibility of achieving an electric control of their magnetic states. The group
of hexaferrites, i.e., iron oxides with hexagonal crystal structures, look promising in view of their high operating temperatures and huge ME effects \cite{Kitagawa10,Kimura12,Chai14,Zhai17} owing to their magnetic structures being very sensitive to the chemical composition and low external magnetic field \cite{Ishiwata08,Kimura12,Wang12}.

\begin{figure*}
	  \centering
	  \includegraphics[width=176mm]{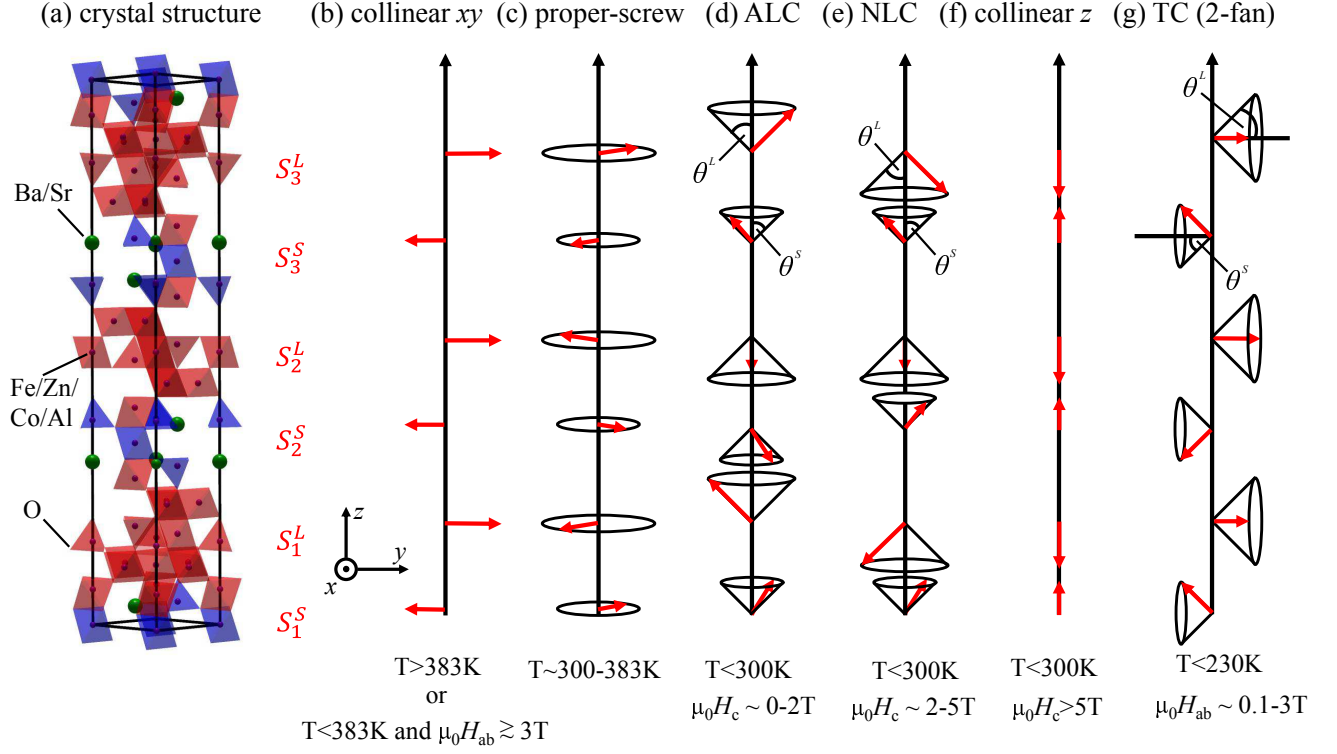}
	  \caption{(a) Crystal and (b)-(g) magnetic structures of the
	  Y-hexaferrite BaSrCoZnFe$_{11}$AlO$_{22}$. Below each magnetic structure, its occurrence in the
	$(H,T)$ phase diagram is marked.}
	  \label{fig:structures}
  \end{figure*}

In terms of crystal structures, hexaferrites are classified as M-, Y-, Z-type
etc.\ depending on the stacking sequence of basic \textit{crystallographic}
blocks along the hexagonal axis \cite{Kimura12,Pullar12}. Owing to their
complexity, the magnetic structures determined by neutron diffraction are
usually described via different, \textit{magnetic} blocks also aligned along the
hexagonal axis, denoted as L and S, possessing large and small magnetic moments,
respectively  \cite{Momozawa85,Ishiwata08,Soda11,Nakajima16a}. Within each
individual block, the magnetic moments are collinearly aligned \cite{Kimura12}.
Hexaferrites mostly possess ferrimagnetic structures; in some cases, the moments
may compensate and yield a zero net magnetic moment \cite{Lee11}, but mostly,
the magnetic structures may easily become ferrimagnetic upon applying moderate
magnetic fields. The magnetic frustration due to the competing superexchange interaction across the boundary between L and S blocks often
yields a noncollinear alignment of spins, leading to the transverse conical (TC)
spin structure [Fig.~\ref{fig:structures}(g)] which induces the electric dipole
moment. This phenomenon can be explained by the inverse Dzyaloshinskii-Moriya
(iDM) interaction \cite{Sergienko06} or Katsura-Nagaosa-Balatsky model
\cite{Katsura07}: $\overrightarrow{P}=\sum\nolimits_{i,j} P_{i,j} \cdot
\overrightarrow{e}_{i,j} \times(\overrightarrow{S}_{i} \times
\overrightarrow{S}_{j})$. In contrast, the dipole moments induced by the iDM
interaction cancel out in the longitudinal conical (LC) spin structures [Figs.~\ref{fig:structures}(d),(e)] where no net polarization is observed.

The first discovered ME hexaferrite was the Y-type
Ba$_{0.5}$Sr$_{1.5}$Zn$_2$Fe$_{11}$O$_{22}$ reported by Kimura \textit{et al.}
\cite{Kimura05}. Later, Wang \textit{et al.} \cite{Wang12} studied a compound
with a similar
composition, BaSrCoZnFe$_{11}$AlO$_{22}$, which we study in this paper,
showing a stronger ME effect at higher temperatures. Near 400\,K, its
paramagnetic structure transforms to a collinear ferrimagnetic one with spins
aligned within the \textit{ab}-plane \cite{Hirose14}. Below $T_{\rm con}\approx
365\,\rm K$, the proper-screw [also called transverse-spiral,
  Fig.~\ref{fig:structures}(c)] magnetic structure is established \cite{Wang12}.
  Upon zero-field cooling (ZFC), the spins start to tilt from the hexagonal
  plane, giving rise to one of the LC structures.

In a part of the Ba$_{x}$Sr$_{2-x}$Zn$_{y}$Co$_{2-y}$Fe$_{11+z}$Al$_{1-z}$O$_{22}$ compounds, two types of LC structures were identified \cite{Lee11,Nakajima16a,Shen17}: (i) The normal longitudinal conical [NLC, Fig.~\ref{fig:structures}(e)], which is stable when the values of $\mathbf{\mu_0}\bm{H}\parallel c$ lie within 2--5\,T, and (ii) the alternating longitudinal conical [ALC,  Fig.~\ref{fig:structures}(d)], which is observed at lower fields ($\mathbf{\mu_0}\bm{H} \lesssim 2\,\rm T$), where the \textit{c}-components of the magnetic moments are aligned as \mbox{\textuparrow-0-\textdownarrow-0} \cite{Nakajima16a} or \mbox{\textuparrow-\textuparrow-\textdownarrow-\textdownarrow} \cite{Shen17}, leading to a zero net magnetic moment. The latter spin configuration gives rise to ferroelectricity even at room temperature due to the magnetostriction (also called exchange-striction) mechanism
$\overrightarrow{P}=\sum\nolimits_{i,j} \overrightarrow{P}_{i,j}
(\overrightarrow{S}_{i} \cdot \overrightarrow{S}_{j})$ \cite{Shen17}. In both
NLC and ALC structures, the spin modulations point along the \textit{c}-axis
with a commensurate ordering of cones---$\bm{Q}_{\rm{C}}=(0,0,1.5)$ for the ALC
structure and $\bm{Q}_{\rm{C}}=(0,0,3)$ for the NLC structure---, and an
incommensurate one $\bm{Q}_{\rm{IC}}=(0,0,q)$ describing the helical angle
dependence. The first-order metamagnetic transition between the NLC and ALC
structures shows a remarkable hysteresis \cite{Nakajima16a,Shen17}. With
increasing $\bm{H}\parallel c$, the conical angle of the NLC structure
decreases and it vanishes at $\mu_0\bm{H}\approx 5\,\rm T$ in
BaSrCoZnFe$_{11}$AlO$_{22}$. Around this value, the magnetization saturates and
the magnetic structure transforms to the collinear ferrimagnetic, as the
magnetic moments of the L and S blocks, pointing along the \textit{c}-axis, become antiparallel\cite{Wang12}. In this case no ferroelectric polarization is induced.

When applying the magnetic field in the magnetically isotropic \textit{ab}
plane, the TC structure with a modulation vector $\bm{Q}_{\rm{C}}=(0,0,1.5)$ is
established \cite{Lee11,Nakajima16a,Shen17}, giving rise to electric
polarization due to the iDM interaction \cite{Shen15}. According to the iDM
term, the polarization is perpendicular to both the \textit{c} axis and the magnetic field. The sign of the polarization is given by the helicity of the spin structure, which is determined by the cross product of electric and magnetic poling fields \cite{Chai14}. When rotating the magnetic field within \textit{ab}-plane, the polarization rotates the same way \cite{Kimura05}. When the magnetic field is reversed in the \textit{ab}-plane, two cases may occur \cite{Chai17}: (i) If the TC phase remains metastable at zero field (case of lower temperatures), the cones' axes rotate in the \textit{ab}- plane conserving the helicity and, subsequently, the polarization is reversed. (ii) If the TC structure is unstable in the low-field region even after \textit{ab}-field poling (at higher temperatures), the cone axes rotate through the \textit{c}-axis (yielding the ALC structure which is stable after ZFC). Consequently, the helicity is reversed, and the polarization is recovered with the same sign after a magnetic field switch \cite{Chai17}.

The dynamic ME effect exhibits resonances called electromagnons. These are electric-dipole-active excitations represented by collective spin motions,
believed to be caused by the same types of microscopic mechanisms as the static ME effect, linked to the ground state magnetic structure. However, electromagnons involve both ground and excited states, thus obeying different selection rules than the static ME effect\cite{Matsumoto17}. It is important to note that electromagnons do not have to be  magnetic-dipole active; i.e., they may represent changes in the magnetic quadrupole (or higher-order multipole) moments where no net magnetization is changed. An analogy is known from lattice dynamics, where some phonons can be non-polar.

The first electromagnon in hexaferrites, reported in
Ba$_2$Mg$_2$Fe$_{12}$O$_{22}$ in the THz range, was attributed to the magnetostriction mechanism \cite{Kida09a}. In
BaSrCo$_2$Fe$_{11}$AlO$_{22}$, a compound very similar to BaSrCoZnFe$_{11}$AlO$_{22}$ investigated here, a similar spin excitation was also observed by inelastic neutron scattering  \cite{Nakajima16a} but not yet by THz spectroscopy which would confirm its electric-dipole activity. Nakajima \textit{et al.} \cite{Nakajima16a} only suggested that the spin excitation could be an electromagnon, since the magnetic structure allows its activation in the THz dielectric spectra via the magnetostriction mechanism. The leading terms contributing to the dynamical electric polarization $\overrightarrow{P}$ are of the type $\overrightarrow{P}\propto \overrightarrow{S}_i\cdot\overrightarrow{\mathrm{\delta}S}_j$, where $\overrightarrow{\mathrm{\delta}S}_j$ represents fluctuations of the neighboring spin j. Since $\overrightarrow{\mathrm{\delta}S}_j$ is perpendicular to $\overrightarrow{S}_j$, the term is higher when the spins are less collinear, in contrast to the static magnetostriction, proportional to $\overrightarrow{S}_i \cdot \overrightarrow{S}_j$, whose effect is the highest in collinear structures. Nakajima \textit{et al.} \cite{Nakajima16,Nakajima16a} proposed that the spin excitations in Ba$_2$Mg$_2$Fe$_{12}$O$_{22}$ and BaSrCo$_2$Fe$_{11}$AlO$_{22}$ correspond to spins oscillations without an influence on the overall magnetic dipole moment; only the quadrupole moment can be changed. Then, the possible electromagnon in BaSrCo$_2$Fe$_{11}$AlO$_{22}$ would be purely electric-dipole active in the THz (or infrared) spectra. Such a coupling between the magnetic quadrupole and the electric dipole moments with the same symmetry can be explained by theory based on symmetry\cite{Wang16,Matsumoto17}.

Here we study ME properties of BaSrCoZnFe$_{11}$AlO$_{22}$ single crystals. We combined different static and dynamic measurements in magnetic field applied both within and perpendicular to the hexagonal plane to explore the $(T,H)$ phase diagram, including the TC, ALC, NLC, transverse-spiral and collinear ferrimagnetic phases. The static measurements include magnetization curves and magnetic-field dependent permittivity; the dynamic properties were probed by THz, infrared, and Raman spectroscopies providing access to excitations with different selection rules. In all measurements, the magnetic field history was recorded, as it may have an influence on the physical properties of the samples. We observed an electromagnon and thoroughly studied its behavior depending on the magnetic field direction and history. The electromagnon strength is clearly correlated with the static magnetic measurements reflecting the magnetic structure, which gives us a tool to check the selection rules activating the electromagnon, including quantitative evaluation. By comparing the analytically calculated electric polarization with the measured electromagnon strength in all observed magnetic phases, we conclude that the electromagnon is most probably caused by the magnetostriction mechanism, as proposed earlier, whereas the spins vibrate along the hexagonal axis.

\section{Samples and experimental details}

${\rm Ba}{\rm Sr}\rm {CoZnFe}_{11}{\rm Al}{\rm O}_{22}$ single crystals were grown by the flux method \cite{Momozawa87}. The samples were measured as grown except for magnetic-field dependent permittivity, before which the samples were annealed in an oxygen atmosphere at 900\degree C for 7 days. The exact composition was determined by EDAX (energy-dispersive analysis of X-rays) as ${\rm Ba}_{1.1}{\rm Sr}_{0.9}{\rm Co}_{1.3}{\rm Zn}_{0.7}{\rm Fe}_{11}{\rm Al}{\rm O}_{22}$.

THz complex transmittance from 0.2 to 2.3\,THz was measured using a custom-made time-domain spectrometer powered by a Ti:sapphire femtosecond laser with 35-fs-long pulses centered at 800\,nm. The system is based on coherent generation and subsequent coherent detection of ultra-short THz transients \cite{Kuzel00}. The detection scheme is realized on an electro-optic sampling of the electric field of the transients within a 1-mm-thick, (110)-oriented ZnTe crystal as a sensor. This allows us to measure time profile of the THz transients transmitted through a studied sample. Details about the calculations of complex index of refraction can be found in Ref. \cite{Kuzel00} Spectra were obtained with resolution better than 0.1 THz. For the low-temperature THz complex transmittance and IR reflectivity spectroscopies, an Oxford Instruments Optistat optical continuous He-flow cryostats with mylar and polyethylene windows, respectively, were used. THz spectroscopy with magnetic field was performed using another custom-made time-domain spectrometer comprising an Oxford Instruments Spectromag cryostat with a superconducting magnet, allowing us to apply an external magnetic field of up to 7\,T in both Voigt and Faraday geometries.

For Raman studies, a Renishaw RM\,1000 Micro-Raman spectrometer equipped with a CCD detector and Bragg filters was used. The experiments were performed using an Ar$^{+}$ ion laser (wavelength of 514.5\,nm) in the backscattering geometry within the 0.3--24\,THz range, in an Oxford Instruments Microstat continuous-flow optical He cryostat. Further, using a Quantum design MPMS and PPMS instruments equipped with the Andeen-Hagerling 2500A high precision capacitance bridge, we carried out measurements of magnetization and of magnetic-field-dependent permittivity, in a temperature interval from 5--400\,K, with a magnetic field of up to 7\,T.

Low-temperature IR reflectivity measurements in the frequency range 1--20\,THz were performed using a Bruker IFS-113v Fourier-transform IR spectrometer equipped with a liquid-He-cooled Si bolometer (1.6\,K) serving as a detector. Room-temperature mid-IR spectra up to 150\,THz were obtained using a pyroelectric deuterated triglycine sulfate detector.

\section{Results and discussion}

\subsection{Magnetization measurements}

Since several neutron diffraction studies of Ba$_{x}$Sr$_{2-x}$Co$_{y}$Zn$_{2-y}$Fe$_{11+z}$Al$_{1-z}$O$_{22}$ Y-type hexaferrites determined their magnetic structures and correlated them with magnetization data \cite{Lee11,Nakajima16a,Shen17}, it is possible to determine the magnetic structures of our sample just from  measurements of magnetic-field dependent magnetization and permittivity.

To verify the presence of conical structures, we measured the
temperature-dependent magnetization for $\bm{H}\parallel c$ and
$\bm{H}\perp c$ \cite{Suppl-Y-hexaferrite}. The phase transition from the
collinear to the proper-screw structure was revealed at 383\,K (Fig.\ S1 in Ref.
\onlinecite{Suppl-Y-hexaferrite}) which is in agreement with previous results,
as the temperature of the phase transition is sensitive to chemical composition
\cite{Hirose14}. After $\bm{H} \perp c$ poling at
$T\approx 10\,\rm K$, the TC structure remains stable up to 230\,K even at
$\bm{H}=0$ (see Ref.\  \onlinecite{Suppl-Y-hexaferrite} and Fig. S1 therein) which is consistent with the previous work of Shen
\textit{et al.} \cite{Shen14}.

\begin{figure}
	  \centering
	  \includegraphics[width=88mm]{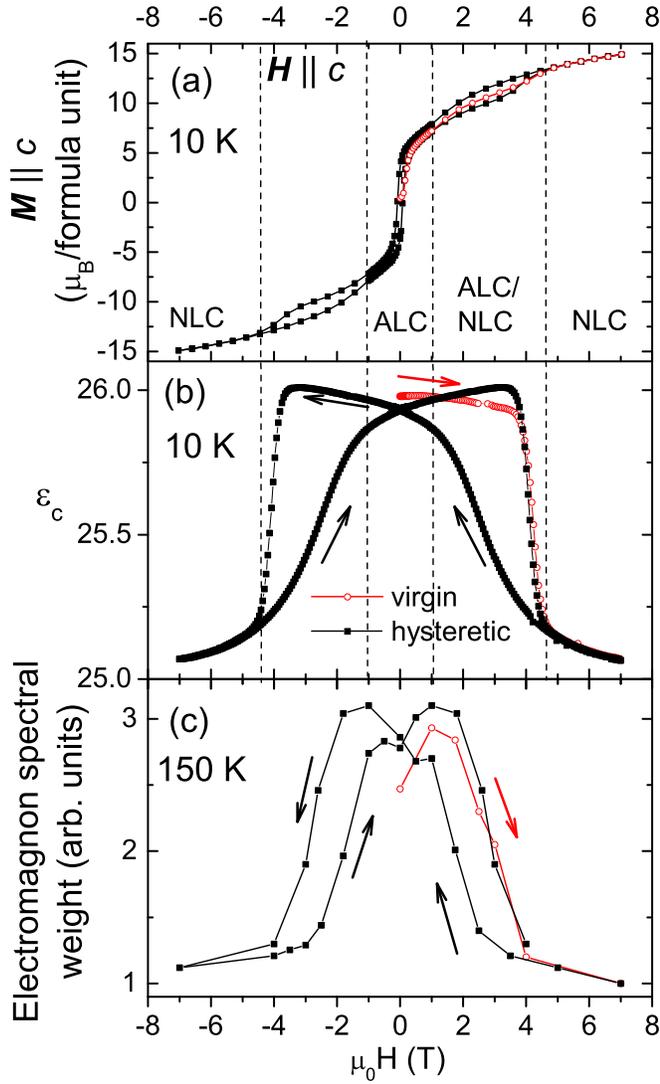}
	  \caption{(a) Magnetization per formula unit, (b) permittivity at 1\,kHz and (c)
	    spectral weight of the electromagnon as a function of
	    $\bm{H}\parallel c$ calculated from THz $k(\omega)$ spectra. Red curves correspond to virgin magnetization
	  curves after ZFC.  Note that at $T=10\,\rm K$ and $T=150\,\rm K$ the phase diagram is qualitatively the same, therefore, we can compare both measurements.} \label{fig:comparison}
  \end{figure}

\begin{figure}[]
	  \centering
	  \includegraphics[width=88mm]{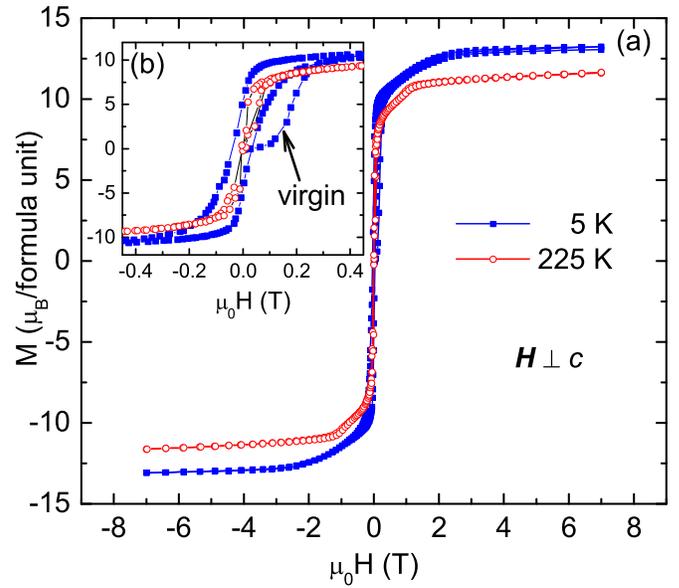}
	  \caption{(a) Magnetization curve for $\bm{H}\parallel$[100]. The inset (b) shows the detailed \textit{M}(\textit{H}) dependence at low fields revealing the magnetization after leaving the virgin state reached by ZFC.}
	  \label{fig:M(H)100}
  \end{figure}

Two magnetic structures, the NLC and ALC one, were reported to exist after ZFC
in Ba$_{x}$Sr$_{2-x}$Co$_{y}$Zn$_{2-y}$Fe$_{11+z}$Al$_{1-z}$O$_{22}$
Y-hexaferrites \cite{Nakajima16,Shen17}. To identify the one present in our
samples at low temperatures, we measured magnetization along the \textit{c}-axis
at 10$\,$K (Fig.~\ref{fig:comparison}(a)). We observed a curve with a remarkable
hysteresis up to $\approx 4\,\rm T$ similar to that observed by Shen \textit{et
al.} \cite{Shen14}, indicating that the ALC structure is the ground state after
ZFC. If the NLC structure were the ground state, the hysteresis would not extend
to such a high magnetic field, since the NLC structure continuously transforms
into the high-field collinear phase. A hysteresis coming from domain switching
would be expected only at low fields, as it is the case of magnetization in the
\textit{ab}-plane (Fig.~\ref{fig:M(H)100}). From the comparison of our
magnetization curve and of the magnetic-field dependent permittivity
(Fig.~\ref{fig:comparison}(a),(b)) with the work of Shen \textit{et al.}
\cite{Shen17}, we can claim that the ALC structure persists up to $\approx
4\,\rm T$ if coming from low-field region, and it recovers only when the field
is decreased to $\approx 1\,\rm T$. A similar behavior was observed at temperatures up to 200$\,$K, so we expect the ALC structure to exist in this temperature region at zero field after ZFC. The existence of ALC structure after ZFC is also supported by THz and Raman measurements (see text below). At higher fields, the NLC structure appears, and it continuously transforms into the collinear ferrimagnetic structure with spins along the \textit{c}-axis.

After applying $\bm{H}$ in the \textit{ab}-plane, the TC structure is known to establish in ${\rm Ba}{\rm Sr}\rm {CoZnFe}_{11}{\rm Al}{\rm O}_{22}$ \cite{Chai14}. In such a case, we observe the virgin magnetization curve outside the hysteresis loop (Fig.~\ref{fig:M(H)100}); this means that after ZFC, the sample has an easy \textit{c}-axis anisotropy, which is consistent with the ALC structure. After a high-field treatment, the zero-field susceptibility is high and it shows only a tiny hysteresis, indicating an easy-plane anisotropy. At $0.3\,\rm T$, the virgin curve coincides with the hysteretic one, which is then the field sufficient to establish the TC structure. This result is consistent with the TC structure persisting also in zero field after applying high $\bm{H}\perp c$, as reported before in \cite{Shen14}.

\subsection{Electromagnon in zero field cooling}

\begin{figure} \centering \includegraphics[width=0.9\columnwidth]{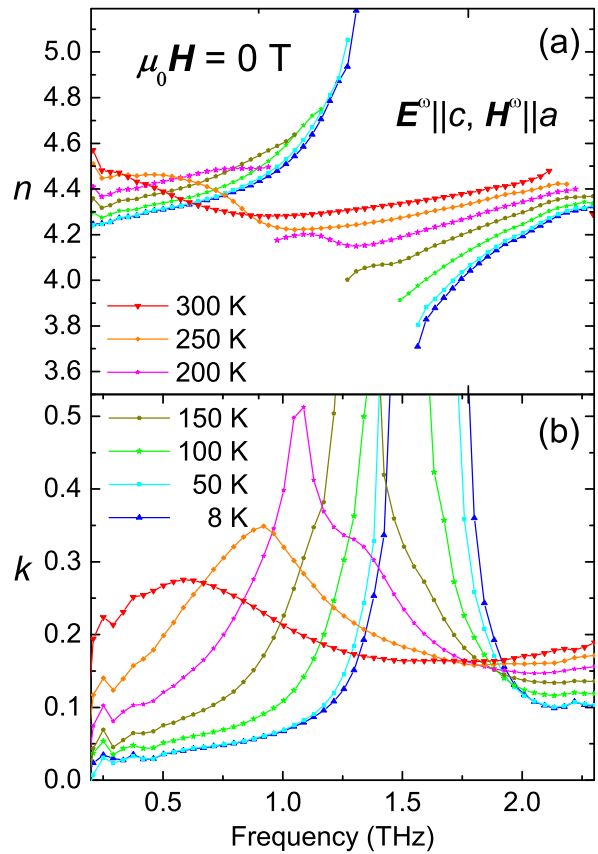}
	\caption{Temperature dependent THz spectra of (a) real and (b) imaginary
	  part of the refractive index in $\bm{E}^{\omega}\parallel c$,
	  $\bm{H}^{\omega}\parallel a$ polarization. Below $200\,\rm K$, the
	parts of the spectra exhibiting high absorption due to the electromagnon are missing.} \label{fig:THz B0} \end{figure}

We measured polarized THz transmittance of ${\rm Ba}{\rm Sr}\rm{CoZnFe}_{11}{\rm
Al}{\rm O}_{22}$ (001)- and (100)-oriented crystal plates, providing a complete
set of spectra  at temperatures from room temperature down to $8\,\rm K$. At
$\approx 1.2\,\rm THz$, we observed an excitation present exclusively in the
$\bm{E}^{\omega}\parallel c$ polarization (Fig.~\ref{fig:THz B0}), implying it
is purely electric-dipole active. If it were magnetically active, it would be
present also in the $\bm{E}^{\omega}\perp c$, $\bm{H}^{\omega}\perp c$ polarized
spectra; however, THz spectra in other polarizations show no remarkable features
(see Fig. S2 and S3 in Suppl. materials \cite{Suppl-Y-hexaferrite}). This polar
excitation is relatively weak and overdamped at room temperature, whereas on
cooling, its damping decreases and its frequency and intensity rise. As we show
below, this excitation strongly depends on magnetic field, therefore, it has a
magnetic origin. Consequently, this is an electromagnon, similar to that
observed in Ba$_2$Mg$_2$Fe$_{12}$O$_{22}$ \cite{Kida09a,Kida11,Nakajima16} and
to a possible one in BaSrCo$_2$Fe$_{11}$AlO$_{22}$ \cite{Nakajima16a}. In
contrast to Ba$_2$Mg$_2$Fe$_{12}$O$_{22}$, where the electromagnon was revealed
below $\approx 100\,\rm K$ \cite{Kida09,Kida11}, we observed the corresponding
absorption up to 300$\,$K. Upon heating, the ALC structure gradually transforms
to the proper-screw one, where the electromagnon is inactive due to the symmetry (see Tab.~\ref{tab:selection rules}); therefore
its strength gradually decreases. At 383\,K, the magnetic structure changes from
the proper-screw to the collinear one, where the electromagnon is also forbidden by
symmetry (see Tab.~\ref{tab:selection rules}).

Investigating the THz spectra in more detail, we see a clear double-peak
structure in the imaginary part of the refractive index $k(\omega)$
(Fig.~\ref{fig:THz B0}b) at $200\,\rm K$. At higher temperatures, the peaks are
also asymmetric. Below $200\,\rm K$, the transmission signal around the peak in $k(\omega)$  is low which prevents us from reliably determining the peak shape \cite{remark1}. The double-peak structure at higher temperatures does not imply the same feature at lower temperatures---it may imply a mixture of magnetic phases, while at low temperatures, usually a single phase is established. The mixed phase was recently reported by Shen \textit{et al.}~\cite{Shen17}, but only after a magnetic field treatment. We therefore believe that the double peak structure persists to low temperatures, and we then see two electromagnon modes with similar frequencies in the pure ALC phase.

\begin{figure}[]
	  \centering
	  \includegraphics[width=88mm]{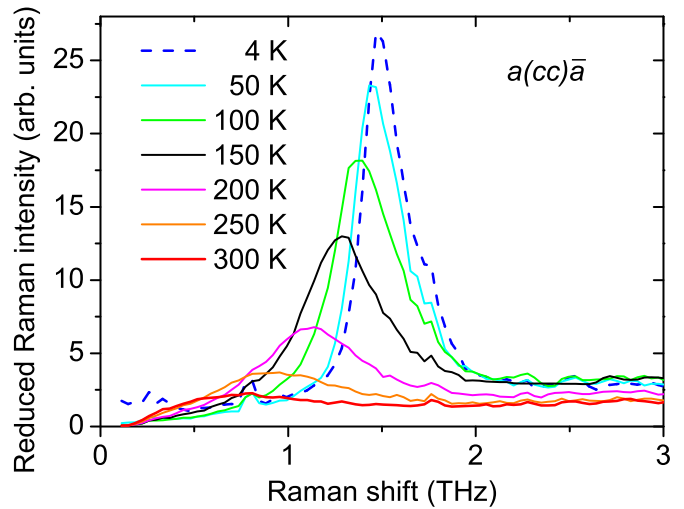}
	  \caption{Temperature dependence of $a(cc)\bar{a}$ Raman spectra
	  showing the electromagnon. The high-frequency phonon
	  spectra are in Fig.~S2 in Ref. \onlinecite{Suppl-Y-hexaferrite}.}
	  \label{fig:Raman}
  \end{figure}

To get a complete knowledge about activity of the electromagnon in various spectra, we measured temperature-dependent polarized Raman spectra (Fig.~\ref{fig:Raman} and Figs. S7, S8 and S9 in Suppl. materials \onlinecite{Suppl-Y-hexaferrite}). In the $a(cc)\bar{a}$-polarized spectra, we clearly see the same electromagnon as in THz spectra. Note also its asymmetric shape confirming its doublet character. Assuming the parent paraelectric space group $D_{3d}^{5}-R\overline3m$,\cite{Wang12} the factor-group analysis of Brillouin-zone-center phonons  reads \cite{Kamba10}
 \begin{eqnarray}
 &\Gamma_{D_{3d}^{5}}=14A_{1g}(a^2+b^2,c^2)+\nonumber \\
 &+4A_{1u}(-)+4A_{2g}(-)+16A_{2u}(c)+\nonumber \\
 &+18E_{g}(a^2-b^2,ab,ac,bc)+20E_{u}(a,b)\,.
 \end{eqnarray}
In such case, the polar $\bm{E}^{\omega}\parallel c$-active electromagnon should
follow the same selection rules as the $A_{2u}$ symmetry polar phonon---it would
be present in the THz spectra but absent in any Raman spectra. Since we see the
electromagnon also in $c^2$-polarized Raman spectra, the magnetic and crystal structures must
be non-centrosymmetric below $\approx 300\,\rm K$ where the electromagnon is observed. It is known that the NLC magnetic phase is centrosymmetric, but the ALC magnetic structure with the \mbox{\textuparrow-\textuparrow-\textdownarrow-\textdownarrow} spin configuration along the $c$ axis breaks the inversion symmetry and, in case of ME coupling, induces electric polarization $\bm{P} \parallel c$ in a low-field region, including $\bm{H}=0$ where this structure exists \cite{Shen17}. Therefore, its point group must be polar $C_{3v}$ and the factor group analysis of phonons reads
\begin{eqnarray}
&\Gamma_{C_{3v}}=30A_{1}(c,a^2+b^2,c^2)+8A_{2}(-)+\nonumber \\
&+38E(a,b,a^2-b^2,ab,ac,bc)\,.
\end{eqnarray}
In this phase, polar phonons and electromagnons have the same $A_{1}$ symmetry and they are active in both \mbox{$\bm{E}^{\omega}\parallel c$}-polarized IR or THz spectra and $c^2$, $a^2$ and $b^2$ Raman spectra, as confirmed by our experiment. (see Fig.~\ref{fig:Raman} and ref. \onlinecite{Suppl-Y-hexaferrite}). The statement that in acentric ferroelectric phases, electromagnons should be both IR and Raman active, was expressed and confirmed first by Skiadopoulou \textit{et al.} for the case of BiFeO$_3$ \cite{Skiadopoulou15}. Nevertheless,  in Raman spectra of BiFeO$_3$ electromagnons are much weaker than phonons \cite{Buhot15}, while in Y-type hexaferrite the electromagnon is stronger than any phonon (Fig. S7 in ref. \onlinecite{Suppl-Y-hexaferrite}). It can be explained by the different mechanism of their activation: Electromagnons in BiFeO$_3$ are induced by iDM interaction, which originates in the spin-orbital coupling which is a weak effect; \cite{Nagel13,Lee16} and the dynamical polarization comes from the electronic polarization \cite{Katsura07} which cannot reach as high values as the ionic one. In contrast, the electromagnon in our sample is activated by magnetostriction (see subsection E below) and this spin-lattice coupling gives the ionic polarization which can be stronger. However more importantly, this ionic, magnetostriction-induced polarization can be a subject of high fluctuations, crucial for high Raman intensity: This comes from the extremely high susceptibility of the frustrated magnetic structures, allowing spin fluctuations with high amplitudes, which is very distinct from the case of BiFeO$_3$. Also the pure magnetic origin is unlikely to be the reason for presence of an electromagnon in Raman spectra \cite{Fleury68a}. A more detailed discussion of the high Raman intensity of the electromagnon can be found in the Supplemental materials \cite{Suppl-Y-hexaferrite}, and the IR and Raman phonon spectra will be presented elsewhere \cite{Kamba18}.

\subsection{Evolution of the electromagnon upon $H\perp c$}

\begin{figure}[!h]
	  \centering
	  \includegraphics[width=88mm]{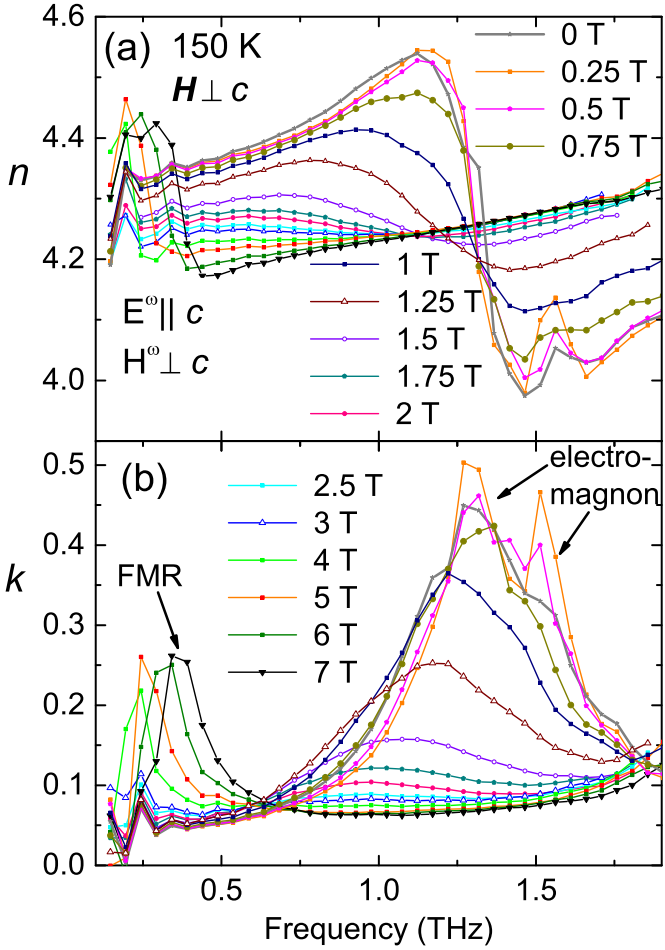}
	  \caption{THz spectra of (a) real and (b) imaginary part of refractive index as a function of external magnetic field $\bm{H}\perp c$ at $150\,\rm K$ with $\bm{E}^{\omega}\parallel c$, $\bm{H}^{\omega}\perp c$. All spectra were taken after applying high magnetic field, therefore, the TC magnetic structure is assumed at low $\bm{H}$.}
	  \label{fig:THz150K B100}
  \end{figure}

Below 270 K, we measured polarized THz spectra in external DC magnetic field,
which stabilizes the TC structure below $\approx 230\,\rm K$.The magnetic-field dependence is qualitatively the same within the whole temperature range; below we discuss the spectra at $150\,\rm K$ where the transmission at the peak position is still above the noise level (Fig.~\ref{fig:THz150K B100}). The strength of the electromagnon gradually decreases with $\bm{H}\perp c $, becoming small above $2\,\rm T$ and negligible above $4\,\rm T$. This corresponds to the continuous phase transition into the collinear ferrimagnetic structure and it is consistent with the assumption that the electromagnon is induced by magnetostriction in the ALC and TC magnetic structures. In the high-field collinear phase, the fluctuations $\overrightarrow{\mathrm{\delta}S}_i$ are perpendicular to all $\overrightarrow{S}_j$, and their scalar products are then practically zero; thus the electromagnon vanishes. Also the gradual decrease in its strength is consistent with the expectations---in the TC structure, for a mode corresponding to spin vibrations along the $c$ axis, its strength should be proportional to the sine of the conical angle $\theta^{S}$ (between the spins and the conical axis in the hexagonal plane; see Tab.~\ref{tab:selection rules} and Fig.~\ref{fig:structures}), and the angle $\theta^{S}$ decreases to zero with increasing magnetic field.

Further, we see a low-frequency resonance which appears at $4\,\rm T$ at
$\approx 0.2\,\rm THz$ and whose frequency increases linearly with the magnetic field with a slope of ca.~$0.05\,\rm THz/T$. Although this resonance reminds of a ferrimagnetic resonance observed in Z-type hexaferrite \cite{Kadlec16}, a detailed analysis reveals a different behavior: First, the slope is much higher than the gyromagnetic ratio for a free electron, $\gamma_0=0.028\,\rm THz/T$, and its value is not a multiple of $\gamma_0$, so the resonance is not likely to be due to a multiple-magnon state. Second, this resonance appears in contradiction with the selection rule for the conventional ferrimagnetic resonance: The resonance should be absent when the oscillating magnetic field $\bm{H}^{\omega}$ is parallel to the spin direction. In our configuration, we set $\bm{H}^{\omega}\parallel\bm{H}$, meaning that some spin directions should be different from $\bm{H}$. However, this is expected at lower fields, while we see the resonance only in the high field region. Moreover, we observed this resonance also in other polarized spectra, which hinders selection rules determination. We believe anyway that it is some kind of ordered magnetic resonance.

\subsection{Evolution of the electromagnon upon $H\parallel c$}
We also measured THz spectra when applying $\bm{H}\parallel c$. Similarly to applying $\bm{H}\perp c$, we observed a suppression of the electromagnon intensity; this occurs at magnetic field values of ca.\ 2$\,\rm$T (see Fig. S4 in Suppl. material \cite{Suppl-Y-hexaferrite}). Furthermore, we also notice a hysteresis of the electromagnon intensity, closely related to that of the magnetization: Fig.~\ref{fig:comparison}(c) shows the magnetic-field dependence of the electromagnon strength (here defined as the integral of $k(\omega)$ over the frequency range of the peak) for $\bm{H}\parallel c$ at $150\,\rm K$. The strength decreases the most at the transition from the ALC to the NLC magnetic structure, and the electromagnon is absent in the saturated state, where the spins are assumed to align collinearly along \textit{c}-direction.

As electromagnons are electric-dipole-active, they contribute to the static
dielectric permittivity. When an electromagnon is suppressed by the magnetic
field, the static permittivity should decrease correspondingly. To verify this,
we measured magnetic-field dependent low-frequency (1\,kHz) dielectric
permittivity $\varepsilon_{c}$ in the \textit{c}-direction since the electromagnon is active for
$\bm{E}^{\omega}\parallel c$. In Fig.~\ref{fig:comparison}(b), we see the
expected decrease in the permittivity at the phase transition from the ALC to
the NLC structure when the electromagnon is suppressed. To evaluate this sum
rule quantitatively, we fitted the THz spectra by the Lorentz oscillator model.
At zero field, the contribution of the electromagnon to the permittivity is 1.4.
Assuming its contribution at 7$\,$T to be zero (verified by the fit), we expect
the same step in the static permittivity; we observed a step of $\approx$1 (Fig.~\ref{fig:comparison}(b)) which is in a rough agreement. The mismatch may come from the conductivity contribution to the permittivity and/or from errors in the size and distance of electrodes of the measured capacitor.

Unfortunately, in Fig.~\ref{fig:comparison} we could not compare the magnetic-field dependent permittivity with the electromagnon spectral weight at the same temperature, because the sample was leaky at 150\,K and the electromagnon absorption was too high at 10\,K (note that we were not able to determine the peak maximum in the $k$ spectrum, because the sample became opaque in this frequency range-see Fig. 4) For that reason the field dependences of the permittivity and of the spectral weight are only qualitatively the same. The disappearance of electromagnons at higher magnetic field explains the decrease in permittivity at 1 kHz with increasing $H$ - see Fig.~\ref{fig:comparison}.

\subsection{Microscopic origin of the electromagnon}

We investigated the electromagnon activity in all principal directions of the DC
magnetic field with respect to the crystallographic axes. We observed the
electromagnon in the TC and ALC magnetic structures, but not in the NLC and
collinear ones (with spins in the \textit{ab}-plane or along the \textit{c}-axis).
Such a comprehensive information enables us to apply to our observations the magnetostriction theory \cite{Kida09,Kida09a,Kida11,Nakajima16,Nakajima16a} describing electromagnons in related Y-type hexaferrites.

Since the \textit{ab}-plane is magnetically isotropic, for describing magnetic states, we can employ a tetragonal basis instead of the hexagonal one; we assume axes \textit{x}, \textit{y}, \textit{z} in the new tetragonal basis coinciding with directions [1,0,0], [-1,2,0], [0,0,1] in the hexagonal system. Our task now is to employ the magnetostriction mechanism possibly inducing the electromagnon to all existing magnetic structures, and to compare the analytically calculated selection rules with the experiment.

The polarization induced by spins according to the magnetostriction model reads
\begin{eqnarray}
\overrightarrow{P}=\sum\limits_{i,j} \overrightarrow{P}_{i,j}(\overrightarrow{S}_{i} \cdot \overrightarrow{S}_{j}) \rm{,}
\end{eqnarray}
\noindent
where the summation involves the nearest neighbors within a magnetic unit cell, which can be quite
large in modulated structures. Taking only nearest neighbors is relevant since spins are quite large in
the block approximation (therefore they can be treated as classical), the next nearest neighbors are far apart from each other, and the superexchange interaction plays the most important role on boundaries of the blocks. The prefactor $\overrightarrow{P}_{i,j}$ must respect the crystal symmetry \cite{Sergienko06b,Wang16}. Taking the Y-hexaferrite crystal structure and the magnetic structure in the block approximation, the polarization along the \textit{z}-axis has the following form \cite{Nakajima16,Nakajima16a,Shen17}:
  \begin{eqnarray}
   P_{z}\propto\sum\limits_{i} (\overrightarrow{S}^{L}_{i} \cdot \overrightarrow{S}^{S}_{i}-\overrightarrow{S}^{S}_{i} \cdot \overrightarrow{S}^{L}_{i+1}) \rm{.}
  \end{eqnarray}
For the dynamic ME effect, we assume all spins in Eq.~(3) as time-dependent,
resulting in a time-dependent polarization $P_{z}(t)$. More specifically, we
formally separate the static equilibrium spins known from the magnetic
structure, and the dynamical part. We assume the dynamical part to be small and
perpendicular to the equilibrium spin direction, as the spin lengths must be conserved. As we are interested only in the dynamic ME effect, we omit the scalar products of static spins and take into account only the first-order dynamic terms of type $\overrightarrow{S}_1 \cdot \overrightarrow{\mathrm{\delta}S}_2$, since the second-order terms of type $\overrightarrow{\mathrm{\delta}S}_1 \cdot \overrightarrow{\mathrm{\delta}S}_2$ are assumed to be small. Next, we separate the contributions to $P_{z}$ coming from \textit{x}, \textit{y} and \textit{z} components of spins' deviations.

We now look for possible modes which can sum up constructively to the
oscillating $P_{z}(t)$. The modes must not be magnetic-dipole-active to be
consistent with the experiment, therefore, they should originate in spin vibration,
not precession. The spin deviations are assumed to be proportional to the original spin lengths $S^{L}$, $S^{S}$. The static magnetic structures are described as follows: Among the TC structures, we take the one called 2-fan, described as \cite{Nakajima16a,Hearmon13}
$$\overrightarrow{S}^{S}_{1}=
\begin{pmatrix}
0\\
-S^{S} \cos{(\theta^{S})}  \\
S^{S} \sin{(\theta^{S})} \\
\end{pmatrix}\rm{,\quad}
\overrightarrow{S}^{L}_{1}=
\begin{pmatrix}
S^{L} \sin{(\theta^{L})}\\
S^{L} \cos{(\theta^{L})} \\
0 \\
\end{pmatrix}\rm{,\quad}$$

$$\overrightarrow{S}^{S}_{2}=
\begin{pmatrix}
0\\
-S^{S} \cos{(\theta^{S})}  \\
-S^{S} \sin{(\theta^{S})} \\
\end{pmatrix}\rm{,\quad}
\overrightarrow{S}^{L}_{2}=
\begin{pmatrix}
-S^{L} \sin{(\theta^{L})}\\
S^{L} \cos{(\theta^{L})} \\
0 \\
\end{pmatrix}$$
where $\theta^{L}$ and $\theta^{S}$ denote the conical angles in the large and
small blocks, respectively, taken from the \textit{y}-axis  (Fig.~\ref{fig:structures}(g)).

For the LC structures, the situation is more complex, since there is also an
incommensurate component, and the magnetic unit cell can be quite large. The
length of the incommensurate modulation vector $\bm{Q}_{\rm{IC}}$ depends on temperature
and it reaches a value of ca.\ 0.7 below 150\,K \cite{Ueda18}. For simplicity, we
use the approximate value of 0.75 which is commensurate; then, the magnetic unit cell is only doubled compared to the TC structure, and contains 8 spins.

For the ALC structure, the spin configuration is the following\cite{Nakajima16a}:
$$\overrightarrow{S}^{S}_{1}=
\begin{pmatrix}
S^{S} \sin{(\theta^{S})}  \cos{(\phi)} \\
S^{S} \sin{(\theta^{S})}  \sin{(\phi)} \\
S^{S} \cos{(\theta^{S})} \\
\end{pmatrix}\rm{,\quad}
\overrightarrow{S}^{L}_{1}=
\begin{pmatrix}
0 \\
-S^{L} \sin{(\theta^{L})} \\
S^{L} \cos{(\theta^{L})} \\
\end{pmatrix}\rm{,\quad}$$

$$\overrightarrow{S}^{S}_{2}=
\begin{pmatrix}
-S^{S} \sin{(\theta^{S})}  \cos{(\phi)} \\
S^{S} \sin{(\theta^{S})}  \sin{(\phi)} \\
-S^{S} \cos{(\theta^{S})} \\
\end{pmatrix}\rm{,\quad}
\overrightarrow{S}^{L}_{2}=
\begin{pmatrix}
S^{L} \sin{(\theta^{L})} \\
0 \\
-S^{L} \cos{(\theta^{L})} \\
\end{pmatrix}\rm{,\quad}$$

$$\overrightarrow{S}^{S}_{3}=
\begin{pmatrix}
-S^{S} \sin{(\theta^{S})}  \cos{(\phi)} \\
-S^{S} \sin{(\theta^{S})}  \sin{(\phi)} \\
S^{S} \cos{(\theta^{S})} \\
\end{pmatrix}\rm{,\quad}
\overrightarrow{S}^{L}_{3}=
\begin{pmatrix}
0 \\
S^{L} \sin{(\theta^{L})} \\
S^{L} \cos{(\theta^{L})} \\
\end{pmatrix}\rm{,\quad}$$

$$\overrightarrow{S}^{S}_{4}=
\begin{pmatrix}
S^{S} \sin{(\theta^{S})}  \cos{(\phi)} \\
-S^{S} \sin{(\theta^{S})}  \sin{(\phi)} \\
-S^{S} \cos{(\theta^{S})} \\
\end{pmatrix}\rm{,\quad}
\overrightarrow{S}^{L}_{4}=
\begin{pmatrix}
-S^{L} \sin{(\theta^{L})} \\
0 \\
-S^{L} \cos{(\theta^{L})} \\
\end{pmatrix}\rm{,\quad}$$
and for the NLC structure:
$$\overrightarrow{S}^{S}_{1}=
\begin{pmatrix}
S^{S} \sin{(\theta^{S})}  \cos{(\phi)} \\
S^{S} \sin{(\theta^{S})}  \sin{(\phi)} \\
S^{S} \cos{(\theta^{S})} \\
\end{pmatrix}\rm{,\quad}
\overrightarrow{S}^{L}_{1}=
\begin{pmatrix}
0 \\
-S^{L} \sin{(\theta^{L})} \\
-S^{L} \cos{(\theta^{L})} \\
\end{pmatrix}\rm{,\quad}$$

$$\overrightarrow{S}^{S}_{2}=
\begin{pmatrix}
-S^{S} \sin{(\theta^{S})}  \cos{(\phi)} \\
S^{S} \sin{(\theta^{S})}  \sin{(\phi)} \\
S^{S} \cos{(\theta^{S})} \\
\end{pmatrix}\rm{,\quad}
\overrightarrow{S}^{L}_{2}=
\begin{pmatrix}
S^{L} \sin{(\theta^{L})} \\
0 \\
-S^{L} \cos{(\theta^{L})} \\
\end{pmatrix}\rm{,\quad}$$

$$\overrightarrow{S}^{S}_{3}=
\begin{pmatrix}
-S^{S} \sin{(\theta^{S})}  \cos{(\phi)} \\
-S^{S} \sin{(\theta^{S})}  \sin{(\phi)} \\
S^{S} \cos{(\theta^{S})} \\
\end{pmatrix}\rm{,\quad}
\overrightarrow{S}^{L}_{3}=
\begin{pmatrix}
0 \\
S^{L} \sin{(\theta^{L})} \\
-S^{L} \cos{(\theta^{L})} \\
\end{pmatrix}\rm{,\quad}$$

$$\overrightarrow{S}^{S}_{4}=
\begin{pmatrix}
S^{S} \sin{(\theta^{S})}  \cos{(\phi)} \\
-S^{S} \sin{(\theta^{S})}  \sin{(\phi)} \\
S^{S} \cos{(\theta^{S})} \\
\end{pmatrix}\rm{,\quad}
\overrightarrow{S}^{L}_{4}=
\begin{pmatrix}
-S^{L} \sin{(\theta^{L})} \\
0 \\
-S^{L} \cos{(\theta^{L})} \\
\end{pmatrix}\rm{. }$$
In both cases, the helical angle $\phi = 45\degree$ for $\bm{Q}_{\rm{IC}}=(0,0,0.75)$, and the relative phase of small and large spin modulations is 180\degree.

The criterion for $P_{z}(t)$ via the magnetostriction is the appearance of oscillating \mbox{\textuparrow-\textuparrow-\textdownarrow-\textdownarrow} spin structure where spins can point in any direction. Taking for example the 2-fan structure and spin components in the \textit{z}-direction, the static magnetic structure is \mbox{\textuparrow-0-\textdownarrow-0}, leading to zero static $P_{z}$. However, if we add small spin deviations with \mbox{\textuparrow-\textuparrow-\textdownarrow-\textdownarrow} amplitudes, we can get oscillating $P_{z}$---this is exactly the so-called out-of-phase mode, proposed by Nakajima \textit{et al.}\cite{Nakajima16}. As the spins oscillate in the \textit{z}-direction, we then call it \textit{z}-mode. In the \textit{xy}-plane, we can find another mode contributing constructively to $P_{z}(t)$, where the neighboring small spins oscillate in opposite directions along \textit{x}-axis and the large spins are not involved; we call this the \textit{x}-mode.  On the other hand, all oscillations in the \textit{y}-direction (the cone axis) add up destructively, yielding no contribution to $P_{z}(t)$. Altogether, in the 2-fan structure, we can expect 2 modes which are listed in Tab.~\ref{tab:selection rules}.

\begin{table*}
  \caption{Analytically calculated activity of electromagnons in
    $\bm{E}^{\omega}\parallel c$ spectra induced by the magnetostriction
    in the Y-type hexaferrite with various magnetic
    structures. $S^{L}$ and $S^{S}$ are the magnitudes of large and small spins,
    respectively; $\theta^{L}$ and $\theta^{S}$ are the conical angles taken
    from the conical axes (\textit{z} in the case of the NLC and ALC structures
    and \textit{y} in the TC structure); $(\mathrm{\delta}S^{j}_{1})_{i}$ denote spin deviations along $i=x$, $y$, $z$ for small and large spin blocks, marked as $j=S$ and $L$, respectively. The proper-screw structure can be taken as a degenerate ALC (or NLC) structure for $\theta^{L}=\theta^{S}=0$, giving no electromagnon mode.}
\begin{ruledtabular}
\begin{tabular}{|c|cr|c|c|}
  Magnetic & \multirow{2}{*}{Oscillating polarization according to Eq. (3): $P_{z}(t)\propto$} & & Distinct & \multirow{2}{*}{Constraints} \\
  structure & & & modes & \\
  \hline
  & $ 4S^{L} \sin{(\theta^{L})} \cdot (\mathrm{\delta}S^{S}_{1})_{x}$ & & \textit{x}-mode & $(\mathrm{\delta}S^{S}_{1})_{x}\propto S^{S}$ \\
  \multirow{1}{*}{2-fan (TC)}& $ +0 \cdot (\mathrm{\delta}S^{S}_{1})_{y}$ & & --- & \\
  & $ +4S^{S} \sin{(\theta^{S})} \cdot (\mathrm{\delta}S^{L}_{1})_{z}$ & & \textit{z}-mode & $(\mathrm{\delta}S^{L}_{1})_{z}\propto S^{L}$ \\
  \hline
  & $2\sqrt{2}\left[S^{L}\sin{(\theta^{L})}\cdot (\mathrm{\delta}S^{S}_{1})_{x}+S^{S}\sin{(\theta^{S})}\cdot (\mathrm{\delta}S^{L}_{1})_{x}\right]$ &\multirow{2}{*}{$\left.\begin{array}{l}\vphantom{I}\\\vphantom{I}\end{array}\right\}$} & \multirow{2}{*}{ \textit{xy}-mode} & \\
    \multirow{1}{*}{ALC}& $ +2\sqrt{2}\left[S^{L}\sin{(\theta^{L})}\cdot (\mathrm{\delta}S^{S}_{1})_{y}+S^{S}\sin{(\theta^{S})}\cdot (\mathrm{\delta}S^{L}_{1})_{y}\right]$ & & & \\
  & $ +4S^{L}\cos{(\theta^{L})}\cdot (\mathrm{\delta}S^{S}_{1})_{z} +4S^{S}\cos{(\theta^{S})}\cdot (\mathrm{\delta}S^{L}_{1})_{z}$ & & \textit{z}-mode & $(\mathrm{\delta}S^{S}_{1})_{z}\propto S^{S}\cdot\sin{(\theta^{S})}$, $(\mathrm{\delta}S^{L}_{1})_{z}\propto S^{L}\cdot\sin{(\theta^{L})}$ \\
  \hline
  & $2\sqrt{2}\left[S^{L}\sin{(\theta^{L})}\cdot (\mathrm{\delta}S^{S}_{1})_{x}+S^{S}\sin{(\theta^{S})}\cdot (\mathrm{\delta}S^{L}_{1})_{x}\right]$ &\multirow{2}{*}{$\left.\begin{array}{l}\vphantom{I}\\\vphantom{I}\end{array}\right\}$}  & \multirow{2}{*}{ \textit{xy}-mode} & \\
  \multirow{1}{*}{NLC}& $+2\sqrt{2}\left[S^{L}\sin{(\theta^{L})}\cdot (\mathrm{\delta}S^{S}_{1})_{y}+S^{S}\sin{(\theta^{S})}\cdot (\mathrm{\delta}S^{L}_{1})_{y}\right]$ & & & \\
  & $ +0\cdot (\mathrm{\delta}S^{S}_{1})_{z} +0\cdot (\mathrm{\delta}S^{L}_{1})_{z}$ & & --- & \\
  \hline
  \multirow{1}{*}{collinear}& 0 & & --- & \\
\end{tabular}
\end{ruledtabular}
\label{tab:selection rules}
\end{table*}

The two LC structures differ just in \textit{z}-spin components. The ALC structure has either a \mbox{\textuparrow-\textuparrow-\textdownarrow-\textdownarrow} \textit{z}-component spin structure yielding even static $P_{z}$  or a \mbox{\textuparrow-0-\textdownarrow-0} structure with $P_{z}=0$. Nevertheless, in both cases, adding \mbox{\textuparrow-\textuparrow-\textdownarrow-\textdownarrow} spin deviations in the \textit{z}-direction can lead to oscillating $P_{z}$ as in the case of the 2-fan structure via the \textit{z}-mode. In contrast, in the NLC structure, \textit{z}-spin components add up always destructively yielding no oscillating $P_{z}$, as demonstrated in Tab.~\ref{tab:selection rules}.

For the two LC structures in \textit{xy}-plane, the oscillations in \textit{x} and \textit{y} directions are equivalent with the same frequency and have to be represented as one mode because of the constraint that the spins deviations must be perpendicular to the original spin direction. This mode summing constructively was proposed by Nakajima \textit{et al.} \cite{Nakajima16a}; it corresponds to clockwise deviations of small spins and counter-clockwise deviations of large spins in \textit{xy}-plane, and we call it here the \textit{xy}-mode.

Altogether, there are two distinct modes contributing to the $P_{z}(t)$ in 2-fan and ALC structures, and one mode in the NLC structure, as summarized in Tab.~\ref{tab:selection rules}.

From the comparison of our analytical calculations and observations, it appears
that the strong absorption in the THz spectra is due to spin oscillations in the
\textit{z}-direction. In fact, we observe the electromagnon in the 2-fan
and ALC phases but not in the NLC one, and the decrease in its strength is the
most pronounced at the phase transition from the ALC to the NLC phase. In
principle, the strong absorption in the ALC structure could be due to the
\textit{xy}-mode, which was proposed by Nakajima \textit{et al.}\  \cite{Nakajima16} for
the related Y-hexaferrite BaSrCo$_2$Fe$_{11}$AlO$_{22}$, and confirmed by
inelastic neutron scattering. Then, the decrease in the electromagnon strength at the
ALC--NLC phase transition would be caused by the spins inclining towards the
\textit{z}-axis as we apply the magnetic field along \textit{z}, because the
spin system would approach the collinear phase. Nevertheless, the abrupt change
in the electromagnon strength supports the \textit{z}-mode as an origin of the
strong electromagnon absorption, because no such change in magnetization was
observed at the corresponding fields [cf.\ Figs.~\ref{fig:comparison}(a) and
(c)].

As the spin deviations are perpendicular to the original spin directions, there
are additional constraints (last column in Tab.~\ref{tab:selection rules}),
providing an insight into the dependence of the oscillating polarization on
conical angles entering into the formulas for the polarization (second column in
Tab.~\ref{tab:selection rules}). First, any deviation is proportional to the
original spin direction, so all terms contain the prefactor $S^{L}S^{S}$. Only
in the case of \textit{z}-mode in the ALC structure, additional constraints
yield nontrivial dependencies of the oscillating polarization on the conical angle:
$P_{z}(t)\propto\cos{(\theta^{L})}\sin{(\theta^{S})}+\cos{(\theta^{S})}\sin{(\theta^{L})}$,
which, for $\theta^{L}=\theta^{S}=\theta$, becomes
$P_{z}\propto\sin{(2\theta)}$, providing the maximum value of electromagnon
absorption at an angle of 45\degree. Such a nontrivial dependence can be the
reason why the electromagnon strength first increases with magnetic field and
starts decreasing only at $\approx 1\,\rm T$ (Fig.~\ref{fig:comparison}(c)).
However, this dependence may be also caused by a more disordered, probably mixed
magnetic structure, leading to a stronger absorption at $\approx 1\,\rm T$.
Finally, the selection rules for the ALC phase could explain the double-peak
structure in ZFC spectra (Fig.~\ref{fig:THz B0})---the \textit{xy} and
\textit{z} modes may appear simultaneously and contribute at different frequencies.

For the 2-fan structure, the predicted dependence of the electromagnon strength on
the conical angle is monotonic, $P_{z}\propto\sin{(\theta)}$. In the
experiment, we see some deviations from this behavior. Surprisingly, at first, the
intensity of the electromagnon increases with magnetic field, reaching a maximum at
$0.25\,\rm T$; only then it starts decreasing (Fig.~\ref{fig:THz150K
B100}). Notice that when reaching $\mu_0\bm{H} \approx 0.25\,\rm T$, a
single-domain 2-fan structure is established. It is worth noting that in BaSrCo$_2$Fe$_{11}$AlO$_{22}$, at similar
fields, Nakajima \textit{et al.}\ observed a similar behavior---a maximum in the
neutron diffraction intensity\cite{Nakajima16a} which was not completely
explained either. Our observation can have three
explanations: First, zero-field magnetic structure was not single-phase TC. Second, the single domain state provides a constructive
interference of polarization in the material and that of the electromagnetic
wave. Third, going beyond the block approximation and
assuming spin directions varying within the blocks, the magnetostriction term may depend
differently on the spin configuration, similarly to the case of the ALC
structure. This would cause the highest
absorption to occur at a general angle between 0\degree{} and 90\degree, as proposed by Kida \textit{et al.}\cite{Kida11}

Let us now comment on the double-peak structure seen in the range from $0\,\rm
T$ to $0.75\,\rm T$ and observed the most clearly at $0.25\,\rm T$
(Fig.~\ref{fig:THz150K B100}(b)). This feature is not seen in the ZFC spectra
plotted in Fig.~\ref{fig:THz B0}; it may be due to different magnetic-field
histories and it may be connected with the 2-fan state, but not with the ALC one. In the 2-fan structure, two electromagnon modes are allowed, which may explain the observed two peaks.

Last but not least, we note that in Table I we listed just purely electrically-active (i.e., vibrational) modes caused by magnetostriction. We neglected the influence of AC magnetic field on these modes, and we did not analyze other possible magnetic-field-active modes (one of them describing possibly the low-frequency resonance seen in Fig. 6 below 0.4 THz). To this aim, we would need to treat the spin Hamiltonian appropriate for this compound which is still an unresolved question, as so far proposed Hamiltonians were not able to describe all magnetic structures observed in Y-hexaferrites.

Finally, it is obvious that not all minor features in spectra can be explained by the simple theory. To this aim, one would have to dispose of a detailed knowledge of the spin configuration which would make it possible to go beyond the block approximation.

\section{\protect Conclusion}

In conclusion, we investigated experimentally static and dynamic magnetoelectric
properties of the Y-type hexaferrite BaSrCoZnFe$_{11}$AlO$_{22}$. Its
magnetic structures were determined by static magnetization measurements, and a purely
electric-dipole active electromagnon was observed by THz and Raman spectroscopies. The Raman intensity of the electromagnon was unusually high. We suggest that this is due to an anomalously high susceptibility of frustrated magnetic structure, but this hypothesis would require deeper theoretical clarification. We also
studied in detail the properties of the electromagnon in various magnetic phases
determined by the magnetic-field direction and history. Using a
 magnetostriction model, it was possible to identify the origin of the
 electromagnon, to explain its magnetic-field dependence and to correlate the electromagnon strength with the static magnetodielectric properties. We described the
dominant features in the field dependence of the spectra, but some minor ones
remain unexplained. In order to gain an even deeper insight into the observed
behavior, more sophisticated theories, going beyond the block approximations,
would be probably required.

\begin{acknowledgments}
The authors thank Jan Prokle\v{s}ka, Vladim\'{i}r Tk\'{a}\v{c}, Petr Proschek, Jan Drahokoupil, Karel Jurek, and \v{S}t\v{e}p\'{a}n Huber for their help with experiments, and Titusz Feh\'{e}r for helpful discussions. This work was supported by the Czech Science Foundation (projects 15-08389S and 18-09265S), the program of the Czech Research Infrastructures, project LM2011025 and  the M\v{S}MT project LD15014. J. V\'{i}t was partially supported by the Grant Agency of the Czech Technical University in Prague No. SGS16/244/OHK4/3T/14. Y. S. Chai, K. Zhai and Y. Sun are supported by the National Natural Science Foundation of China (Grants Nos. 11534015, 11674384, and 11675255).
\end{acknowledgments}

\bibliographystyle{apsrev}

\begin{thebibliography}{38}
\bibitem[{rem()}]{remark1}
\bibinfo{note}{In principle, we could have made the sample thinner and resolve
  the spectra with high absorption. However, we possess just one exemplar of
  the crystal and plan other measurements with it, so we decided to keep the
  crystal as a whole.}
\end{thebibliography}

\clearpage

\renewcommand\thefigure{S\arabic{figure}}

\setcounter{figure}{0}
\setcounter{section}{0}

\maketitle {\Large SUPPLEMENTARY MATERIALS}

\maketitle \section{Magnetization measurements}

\subsection{Magnetization measurements}
\begin{figure}[b]
	  \centering
	  \includegraphics[width=\columnwidth]{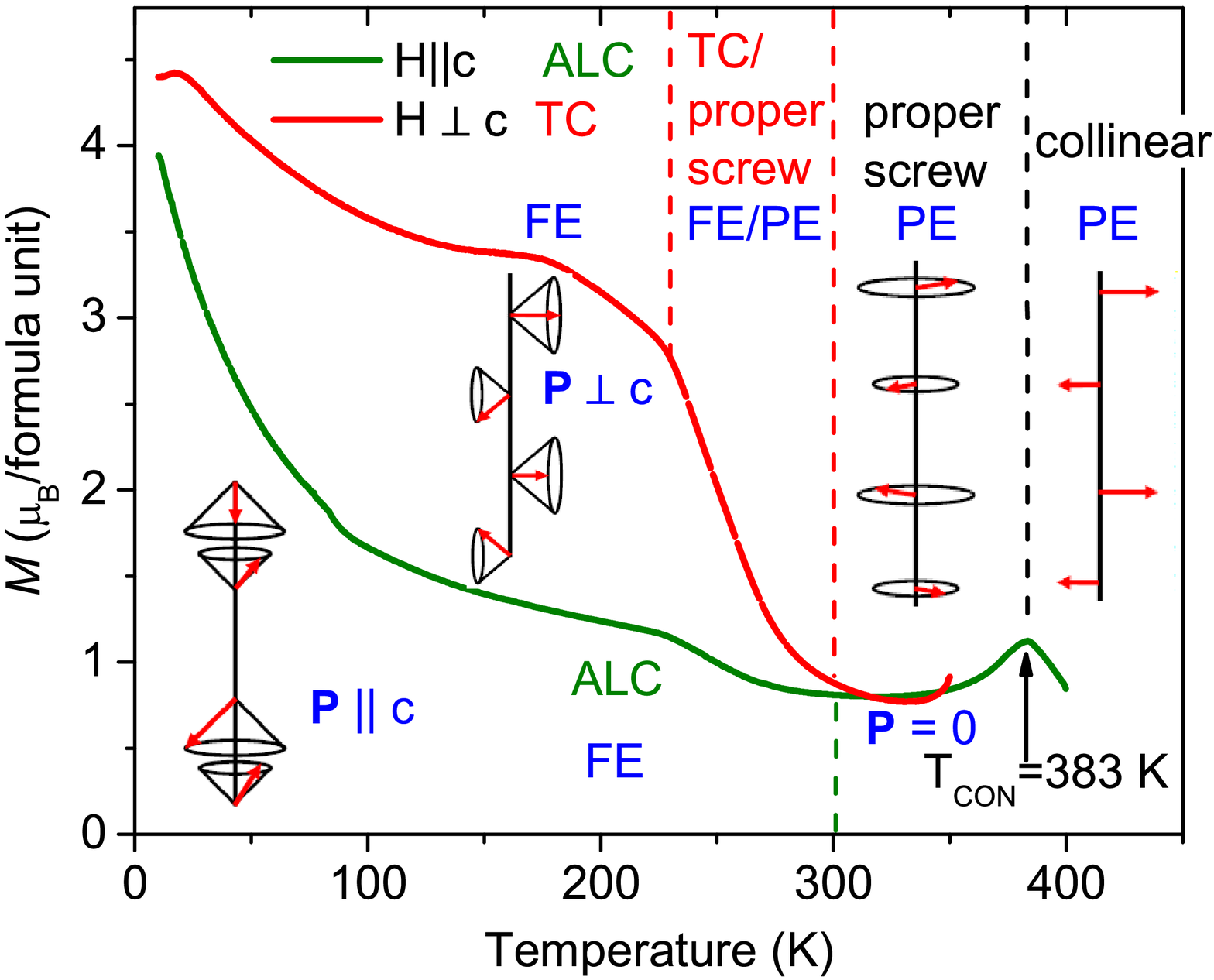}
	  \caption{Temperature dependence of magnetization measured on heating at
	    $\mathbf{\mu_0}\bm{H} = 0.02$\,T after poling at 7\,T at the lowest
	  temperature. The magnetic field was applied $\parallel c$ and $\perp c$
	  directions.}
	  \label{fig:M(T)}
  \end{figure}
 The magnetization curve measured with magnetic field $\bm{H}\parallel c$ (see
 Fig. ~\ref{fig:M(T)}) exhibits a peak at 383\,K, indicating the phase
 transition from the high-temperature collinear phase (with spins in the
 \textit{ab}-plane)  to the proper-screw phase. Wang \textit{et al.} \cite{Wang12} reported a somewhat lower value of $T_{\rm con} = 365\,\rm K$ in ${\rm Ba}{\rm Sr}\rm {CoZnFe}_{11}{\rm
   Al}{\rm O}_{22}$, which can be explained by a slightly different ratio of the Co
   and Zn contents \cite{Hirose14}. The exact composition of our crystal was
   determined by EDAX as ${\rm Ba}_{1.1}{\rm Sr}_{0.9}{\rm Co}_{1.3}{\rm
   Zn}_{0.7}{\rm Fe}_{11}{\rm Al}{\rm O}_{22}$. We note that the $M(T)$ curve
   for our compound is qualitatively similar to that presented by Shen \textit{et al.}
   \cite{Shen14}. Higher values of magnetization at low temperatures may refer to a partial presence of NLC phase after poling at 7 T at the lowest temperature. However, we believe that after zero field cooling, the sample is in the pure ALC state. In the case of $\bm{H}\perp c$, we observed a decrease
   in $\bm{M}$ upon heating from 230\,K to 300\,K. This means that after magnetic-field
   poling at low temperatures, the TC structure is stabilized when heating up to
   230\,K. The same behavior was observed by Shen \textit{et al.} \cite{Shen14}.
   Between 230\,K and 300\,K, a mixture of TC and proper screw phase occurs; above room temperature,
   the proper-screw structure is stabilized irrespective of magnetic field history.

\begin{figure}[!ht] \centering \includegraphics[width=0.9\columnwidth]{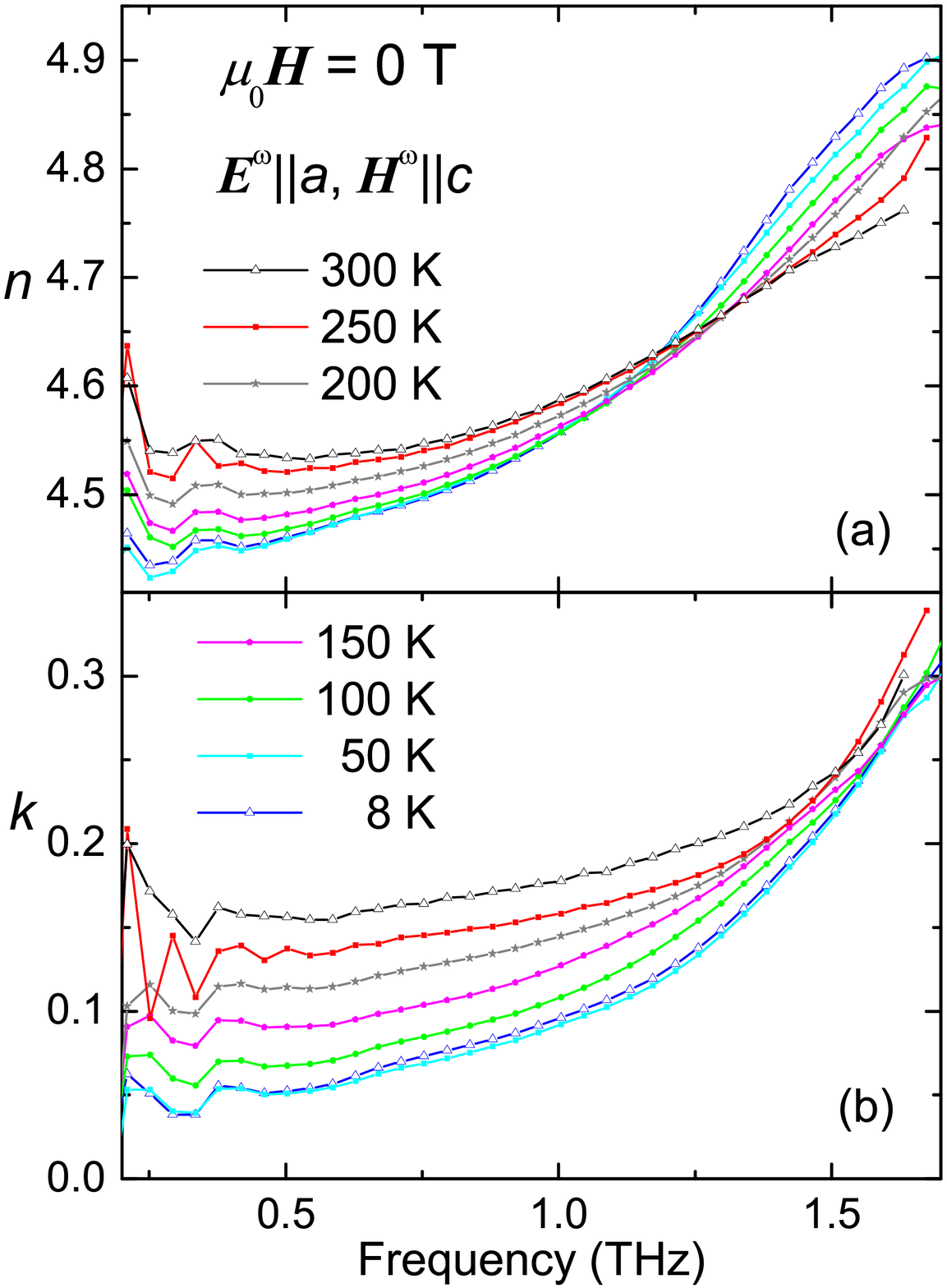}
	\caption{Temperature dependent THz spectra of (a) real and (b) imaginary
	  parts of refractive index in the $\bm{H}^{\omega}\parallel c$, $\bm{E}^{\omega}\parallel a$ polarization. } \label{fig:THz EaHc B0} \end{figure}

\section{Spectroscopic measurements}

\subsection{THz spectra}

\begin{figure}[!ht] \centering \includegraphics[width=\columnwidth]{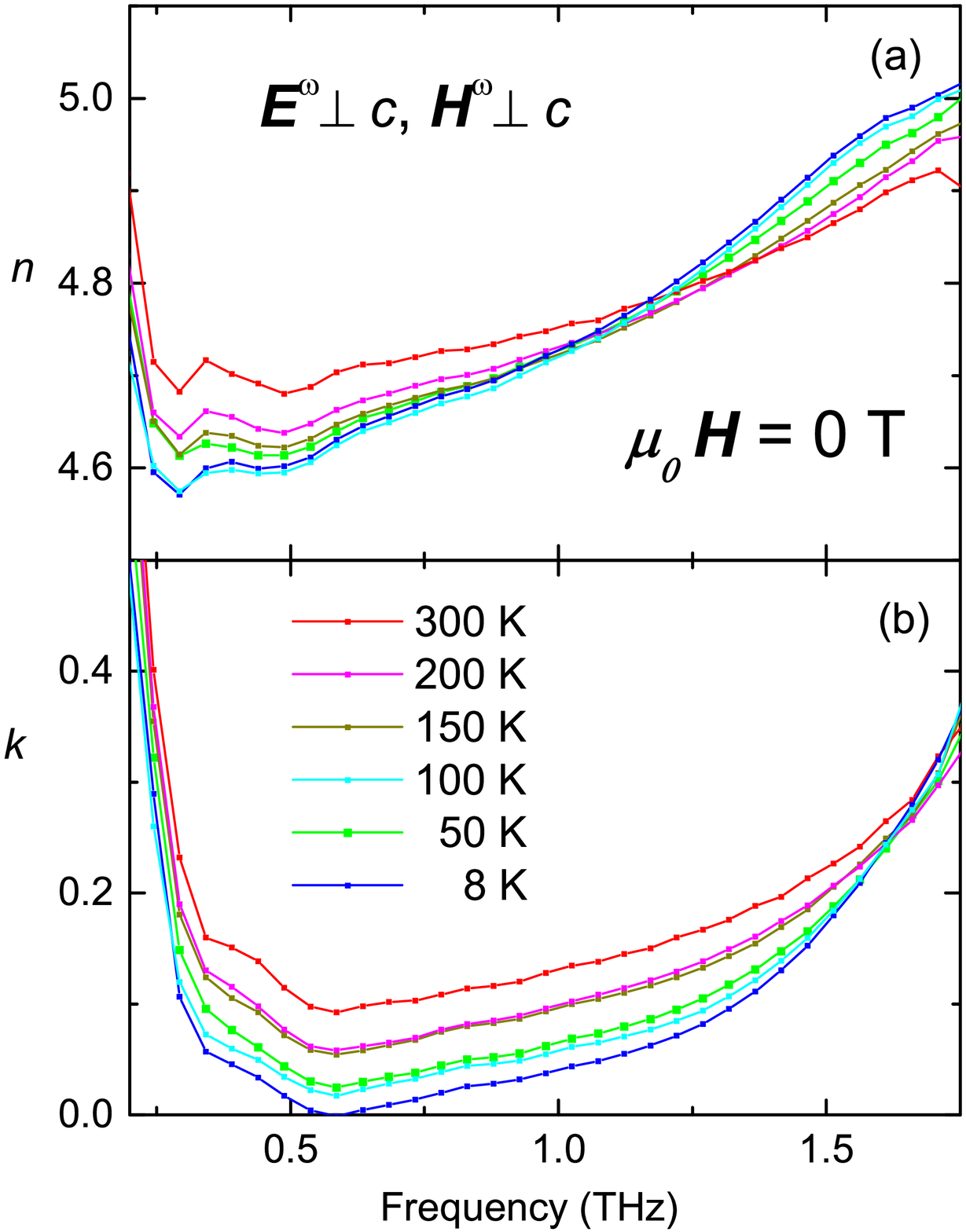}
	\caption{Temperature dependent THz spectra of (a) real and (b) imaginary
	  parts of the refractive index in the $\bm{E}^{\omega}\perp c$, $\bm{H}^{\omega}\perp c$  polarization. The spectra contain no
	spin excitation, only a broad absorption from the phonon near 2\,THz can be seen.} \label{fig:THz EaHb B0} \end{figure}

In contrast to the $\bm{E}^{\omega}\parallel c$ spectra presented in Fig.~4 of
the main text, no spin excitation was observed in the polarized THz spectra
with $\bm{H}^{\omega}\parallel c$,
$\bm{E}^{\omega}\parallel a$ (Fig.~\ref{fig:THz EaHc B0}); here, only a
low-frequency tail from a phonon absorption is present. Also in the
polarized spectra with  $\bm{E}^{\omega}\perp c$, $\bm{H}^{\omega}\perp c$ no
remarkable feature was detected (Fig.~\ref{fig:THz EaHb B0}). This confirms that the spin
excitation is present only in the $\bm{E}^{\omega}\parallel c$-polarized spectra
and therefore it must be an electromagnon, which contributes only to the c-axis dielectric permittivity $\varepsilon_c$.

\begin{figure}[!ht]
	  \centering
	  \includegraphics[width=88mm]{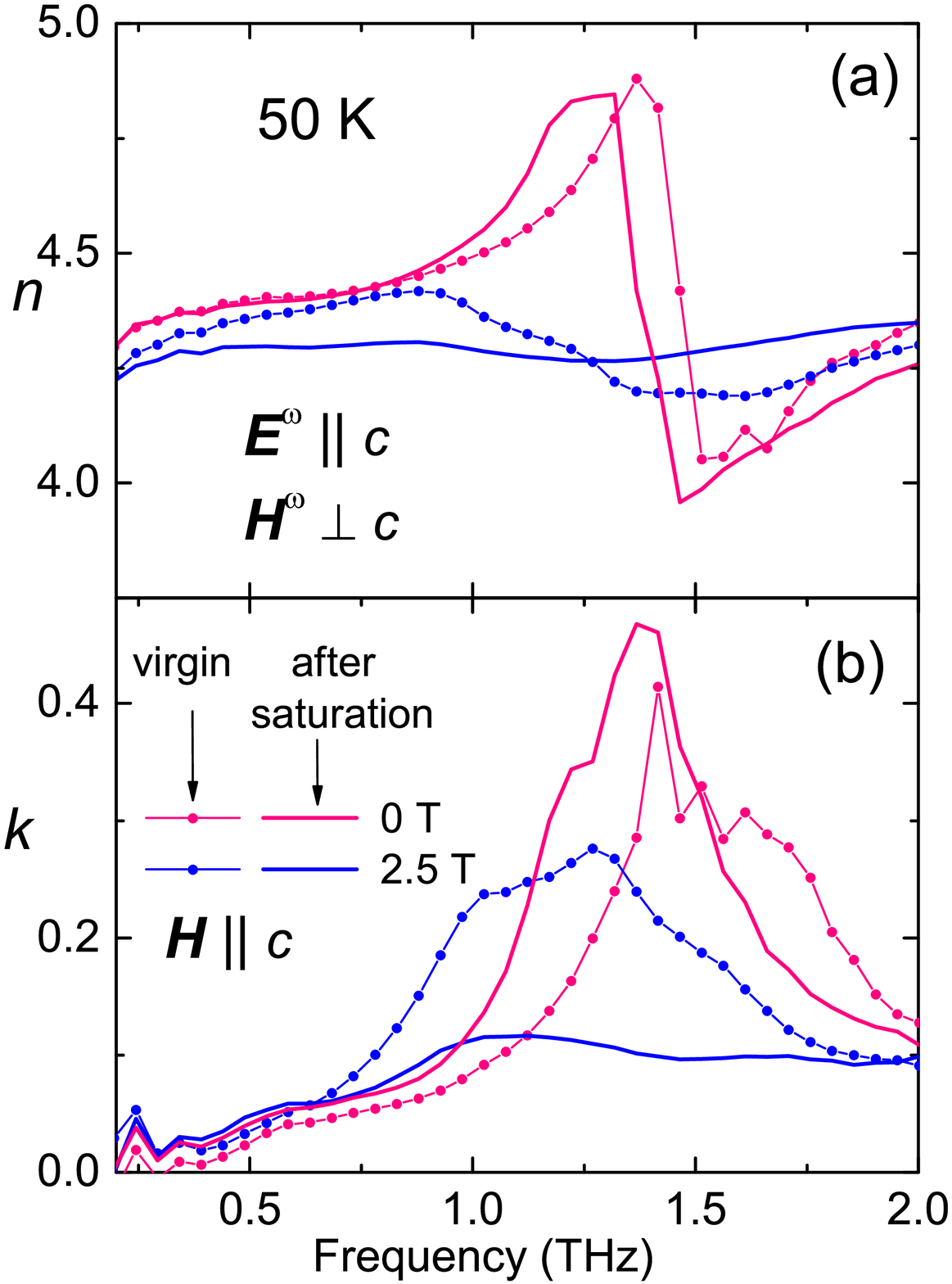}
	  \caption{$\bm{E}^{\omega}\parallel c$-polarized THz spectra of (a)
	real and (b) imaginary parts of refractive index measured at $\mu_{0} H=0$ and
      2.5\,T, before and after applying a magnetic field of 7\,T along [001]. The spectra
    were measured at 50\,K and their shapes are strongly dependent on the magnetization history.}
	  \label{fig:THz50K B001}
  \end{figure}

Fig.~\ref{fig:THz50K B001} shows the complex refractive index spectra with the feature corresponding to the electromagnon, measured at
50\,K with magnetic field values
of 0 and 2.5\,T (applied along the $c$ axis) with different magnetization
history. Clearly, the spectra exhibit substantially different shapes before and
after applying the magnetic field of 7\,T. Note that the magnetization history
influences not only the strength but also the frequency of the electromagnon.

\subsection{Infrared spectra}

\begin{figure}[!ht]
	  \centering
	  \includegraphics[width=\columnwidth]{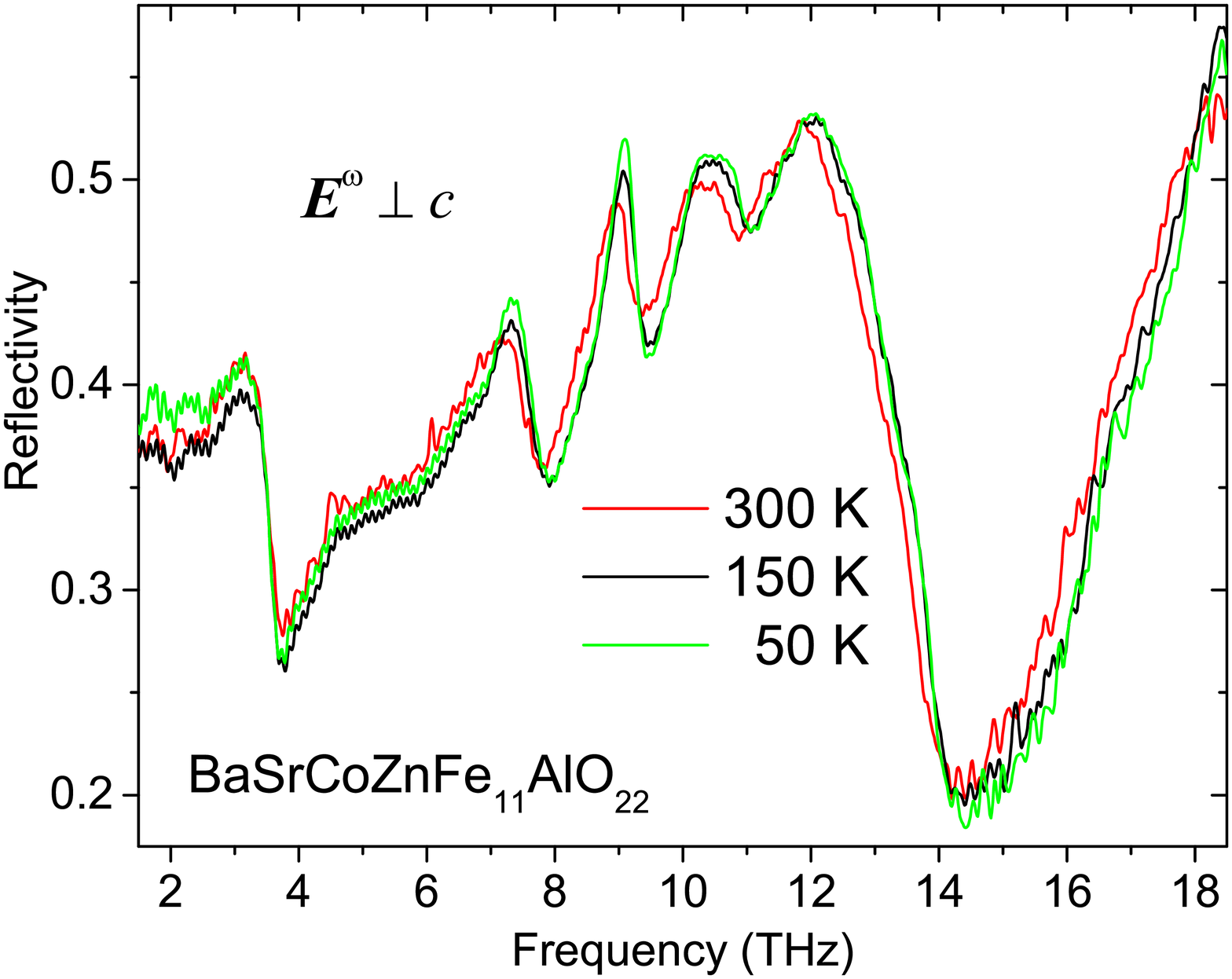}
	  \caption{Far infrared reflectivity spectra at different temperatures for $\bm{E}^{\omega}\parallel a$, $\bm{H}^{\omega}\parallel c$ polarization.}
	  \label{fig:FIR Hc}
\end{figure}

\begin{figure}[!ht]
	  \centering
      \includegraphics[width=\columnwidth]{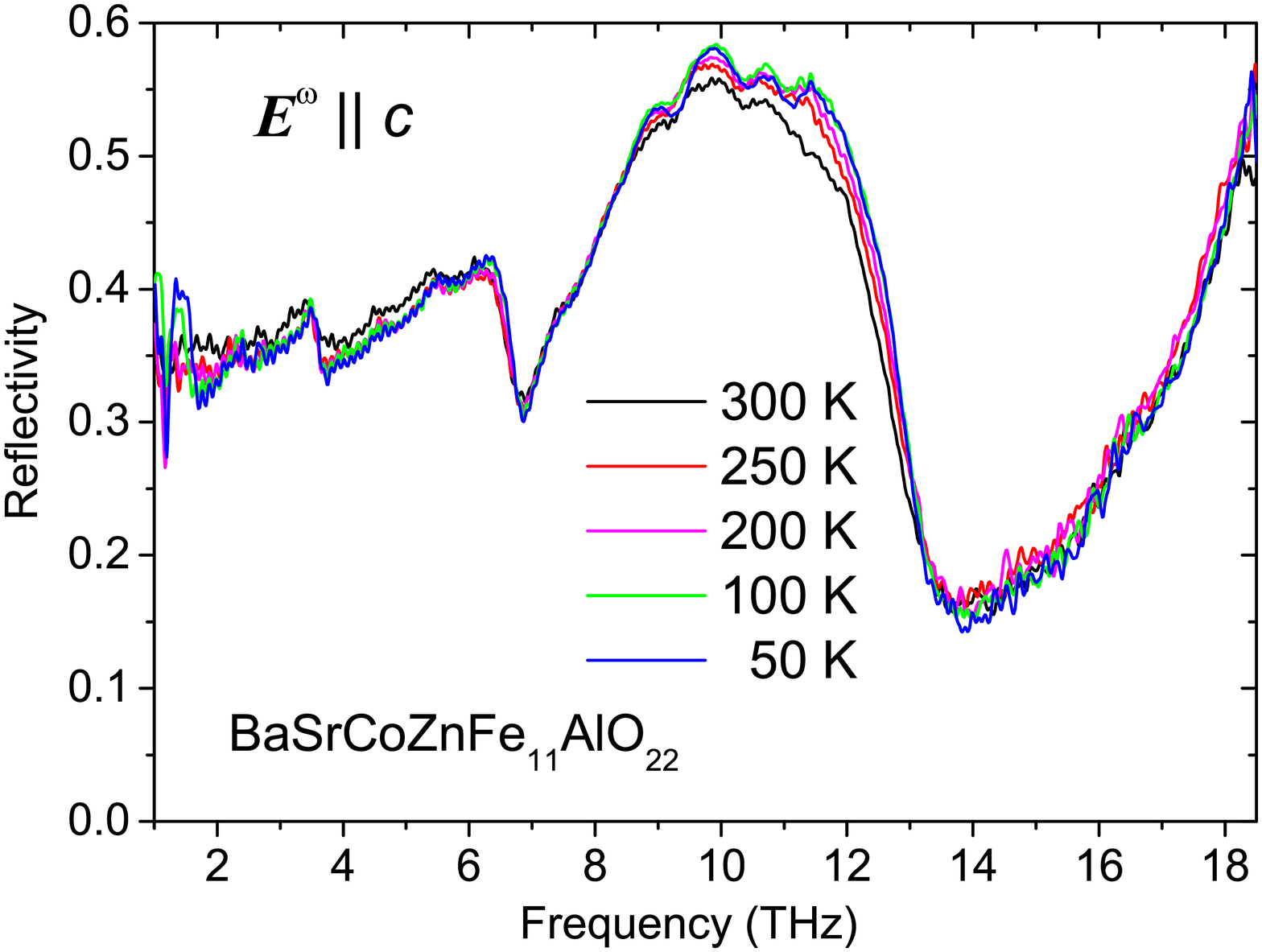}
	  \caption{Far infrared reflectivity spectra at different temperatures for
	    the $\bm{E}^{\omega}\parallel c$, $\bm{H}^{\omega}\parallel a$ polarization.}
	  \label{fig:FIR Ec}
\end{figure}

Figs.~\ref{fig:FIR Hc} and ~\ref{fig:FIR Ec} show the
far-infrared reflectivity spectra in the $\bm{E}^{\omega}\perp c$,
$\bm{H}^{\omega}\parallel c$ and $\bm{E}^{\omega}\parallel c$, $\bm{H}^{\omega}\perp c$ polarizations, respectively, at selected temperatures. As the selection rules for infrared and
THz spectroscopies are the same, the electromagnon should be seen in both kinds
of spectra in $\bm{E}^{\omega}\parallel c$ polarization. Indeed, on cooling, a reflection band arises near 1.5\,THz (Fig.~\ref{fig:FIR Ec}), but
the infrared signal is rather noisy in this range. Consequently, the electromagnon was
evaluated mainly from the THz transmission spectra (see the main text). The phonons
with frequencies above 3\,THz get narrower on cooling, but their number can by explained by selection rules within paraelectric phase with $D_{3d}^{5}$ space group and no
splitting is observed on cooling, so there is no evidence of any ferroelectric distortion in the ALC
magnetic structure. In the other polarization (Fig.~\ref{fig:FIR Hc}), the electromagnon is not seen and the phonons also exhibit only sharpening on cooling.

\subsection{Raman spectra}

\begin{figure}[t]
	  \centering
	  \includegraphics[width=\columnwidth]{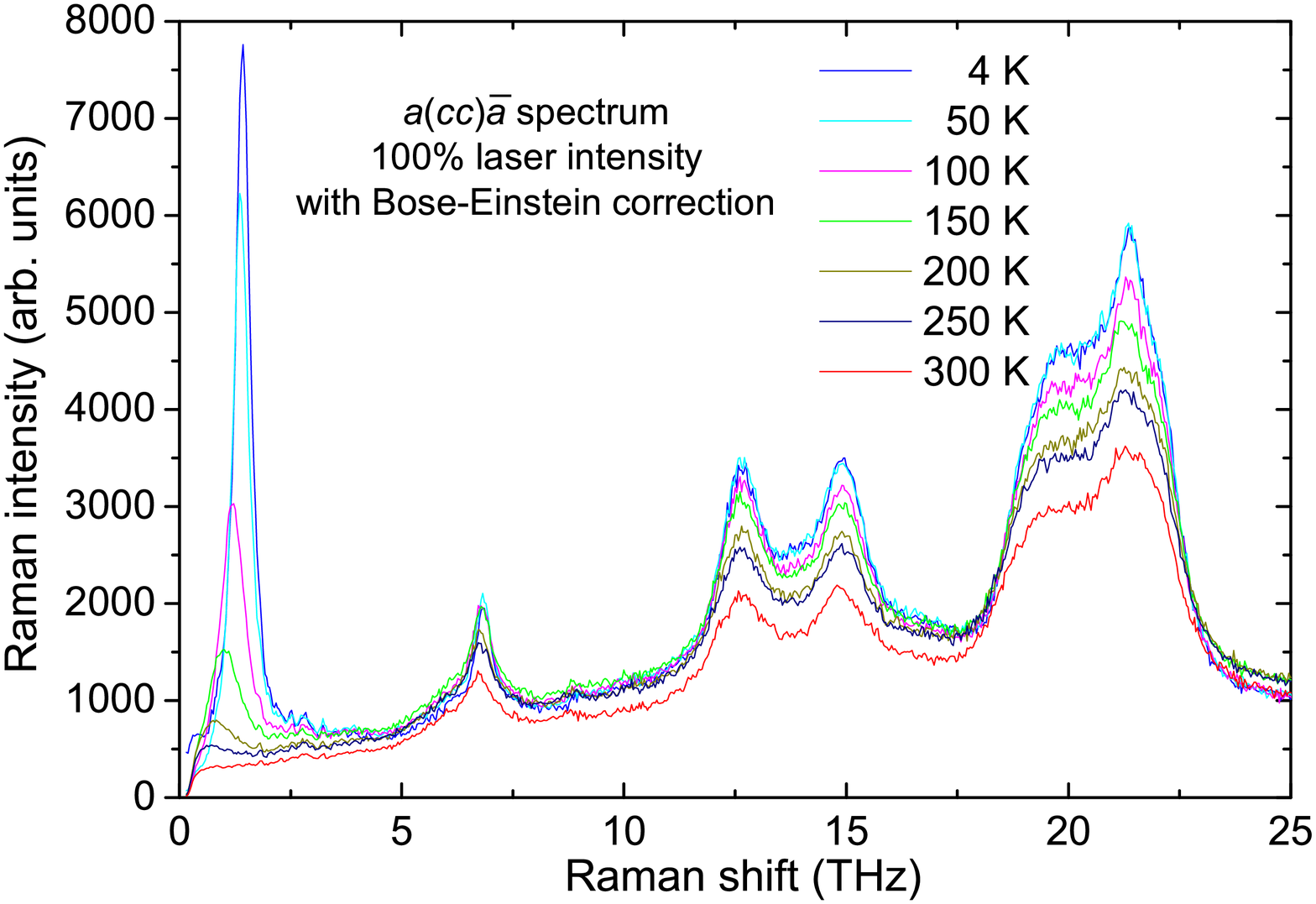}
	  \caption{Temperature dependence of $a(cc)\bar{a}$ Raman spectra
	showing the electromagnon (sharp peak on the left) and phonons at higher
      frequencies. The electromagnon strength is much higher than these of the phonons.}
	  \label{fig:Raman acca}
\end{figure}

Fig.~\ref{fig:Raman acca} shows $a(cc)\bar{a}$-polarized Raman spectra in a
frequency region up to 25\,THz, broader than that presented in Fig.\,5 of the
main text. The electromagnon near 1.5\,THz is much stronger than phonons, suggesting an electric polarization character of this excitation. (See the next section for detailed discussion.)

\begin{figure}
	  \centering
	  \includegraphics[width=90mm]{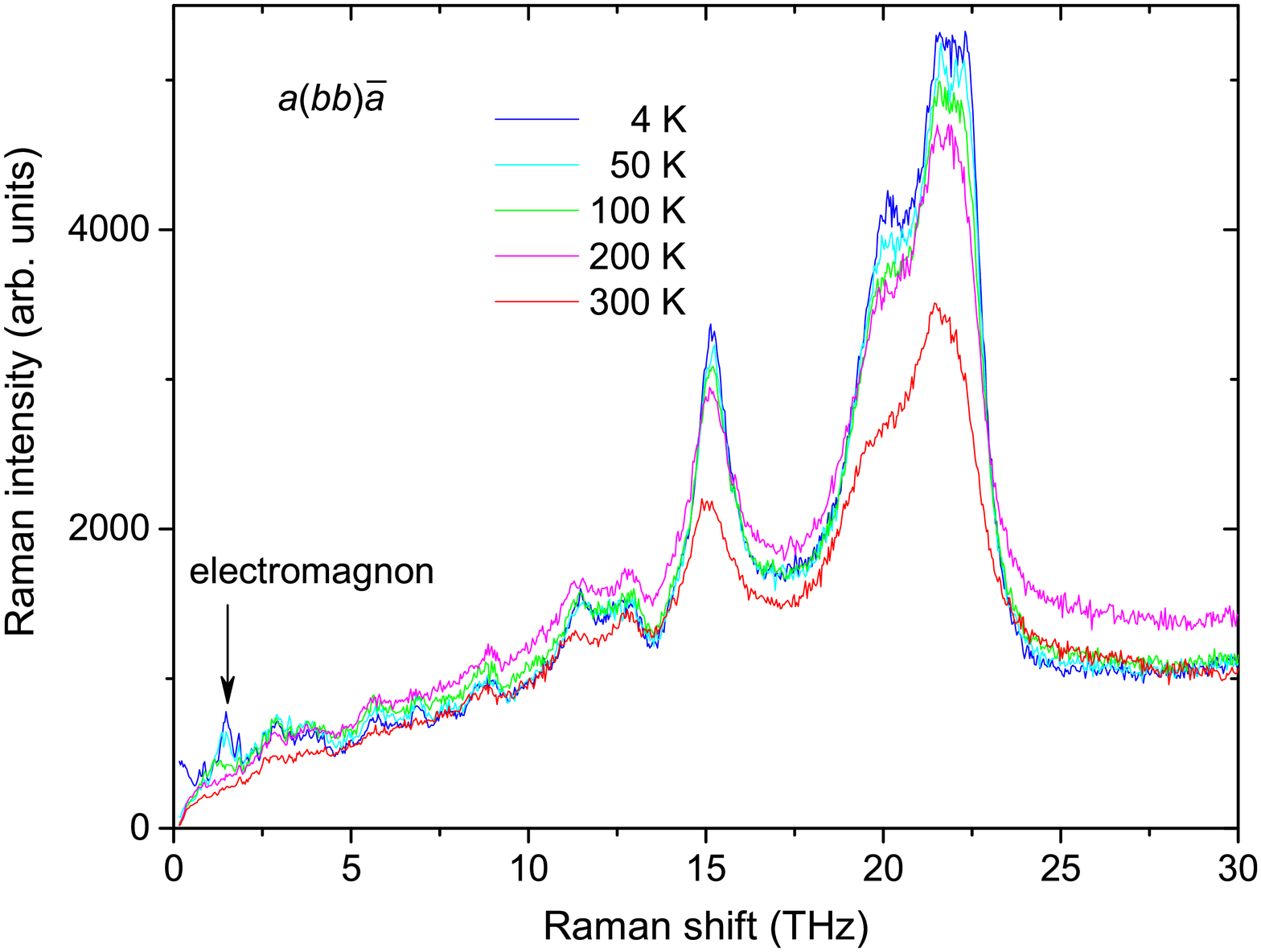}
	  \caption{Temperature dependence of $a(bb)\bar{a}$ Raman spectra.}
	  \label{fig:Raman abba}
\end{figure}

According to the factor group analysis [see Eq.~(2) in the main text], in the polar $C_{3v}$ space group, the electromagnon with an $A_{1}$ symmetry should be also observable in the
$a(bb)\bar{a}$ polarization. We measured also these spectra (see Fig.~\ref{fig:Raman abba}). Here, the intensity of the electromagnon is much lower, nevertheless, we observed the same temperature dependence as in the $a(cc)\bar{a}$ Raman spectra, i.e. the electromagnon frequency hardens on cooling and its damping simultaneously decreases.

\begin{figure}
	  \centering
	  \includegraphics[width=90mm]{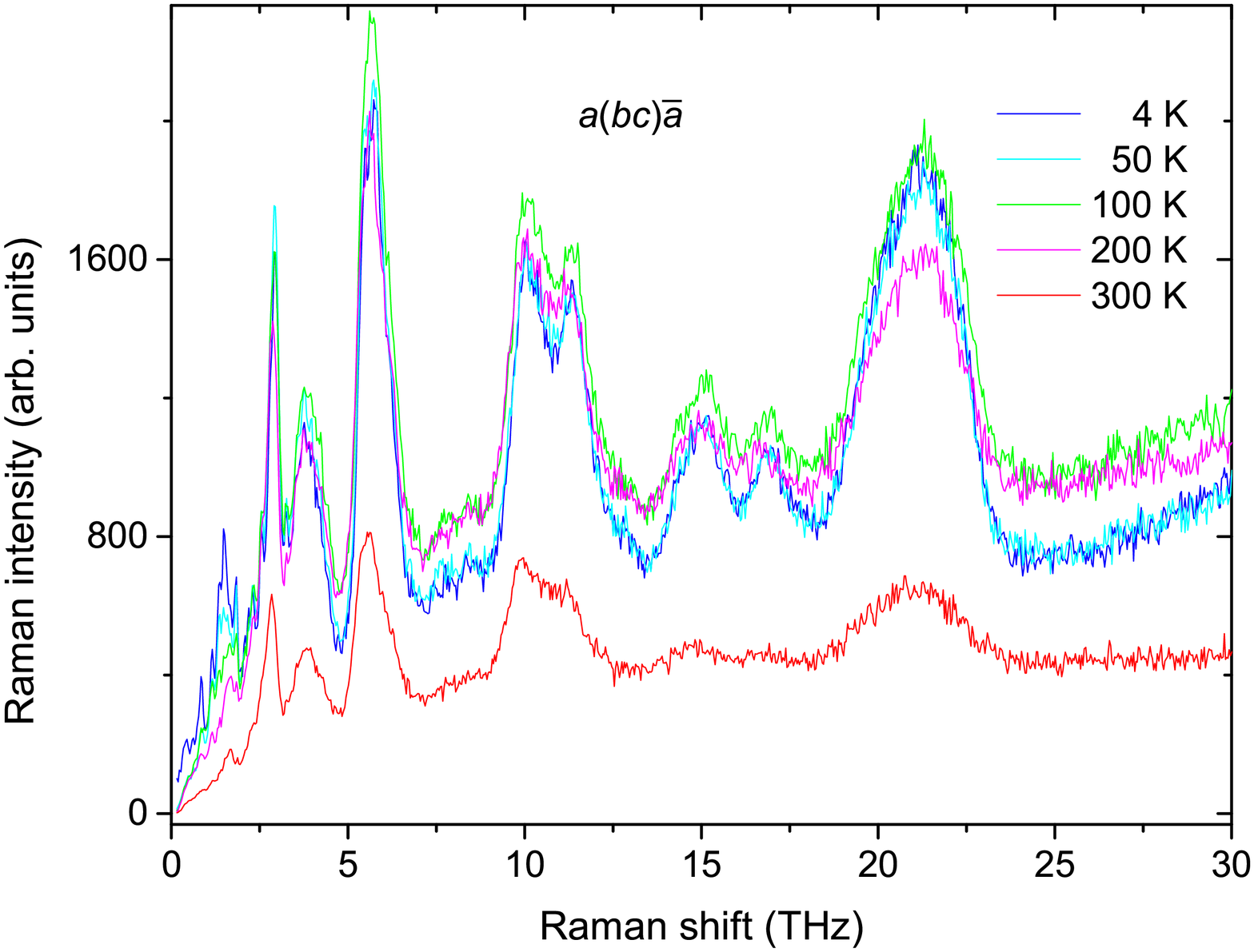}
	  \caption{Temperature dependence of $a(bc)\bar{a}$ Raman spectra.}
	  \label{fig:Raman abca}
\end{figure}

The factor group analysis [see Eq.~(2) in the main text] predicts the
excitations with $E$ symmetry to be active in the $a(bc)\bar{a}$ polarization.
Thus, the electromagnon is expected to be absent from these spectra. Although we
see some absorption peak close to the electromagnon frequency of 1.5\,THz
(Fig.~\ref{fig:Raman abca}), its temperature dependence is distinct from that of
the
electromagnon. In fact, this peak is sharp even at 300\,K and on cooling, the
excitation frequency slightly decreases. Consequently, we attribute this peak to an
$E$ symmetry phonon, similarly to the remaining excitations in the spectra. No phonons were
detected above 25\,THz. We attribute the overall increase in Raman intensity
above 25\,THz to luminescence.

\section{Discussion of IR and Raman activity of the phonons and the electromagnon}

\subsection{Why are there no new phonons after symmetry breaking?}

It is well known that the intensities of the new phonons activated in IR and
Raman spectra below ferroelectric (and structural) phase transitions are, in
general, proportional to some positive integer power of the order parameter
$\eta$. In our case of a spin-induced (i.e., pseudoproper)
ferroelectric phase transition, the magnetic order parameter is related to the
ALC magnetic structure and the ferroelectric distortion arises below $T_{\rm c}$
due to linear coupling of $\eta$ with polarization $P$. In spin-order induced
ferroelectrics, the polar distortion is always very small; therefore, it cannot
be revealed by diffraction studies and mostly only a centrosymmetric space group
of the paraelectric phase is identified. This is also the case of our Y-type hexaferrite \cite{Wang12}.
The phonons newly activated in the ferroelectric phase are often hardly resolved
or even unresolved in the spectra, because they are usually much weaker
than those observed in the paraelectric phase. In the present hexaferrite,
we see no new phonons in the ferroelectric phase; only one new mode
corresponding to the electromagnon appears on cooling in IR and Raman spectra.
This mode is well resolved in THz transmission (Fig. 4 in the main text) and IR
reflectivity (Fig.~\ref{fig:FIR Ec}) and it is dominating the Raman
spectra (Fig.~\ref{fig:Raman acca}). The number of
IR-active phonons observed in our spectra (10 in $\bm{E}^{\omega}\parallel c$ and 7 in
$\bm{E}^{\omega}\perp c$ spectra) can be explained within a non-polar
$D_{3d}^{5}$ space group which does not allow a ferroelectric distortion.
Absence of polar phonons allowed in spin-order-induced ferroelectric structure gives evidence about very small ferroelectric distortion of the crystal lattice and about negligible intensities of possible new modes. The same discussion is valid also for Raman spectra. All observed Raman-active phonons can be explained within the non-polar $D_{3d}^{5}$ space group.

\subsection{Exclusion of the pure magnetic origin of the electromagnon}

An electromagnon intensity so high as in our Raman spectra has never been observed before. This observation lead us to a careful investigation about the reason of such high intensity. First, we excluded a purely magnetic origin of its appearance in the Raman spectra for the following reasons: The Raman intensity is proportional to the polarizability (fluctuations of the refractive index) at the frequency of the laser, which is in visible range. As magnetic excitations lay usually at frequencies well below those of the visible light (the magnetic permeability at frequencies of visible light is roughly 1), the main part of the Raman intensity is given by the electric polarizability (fluctuations of permittivity) in visible range. The direct magnetic-dipole coupling is then usually very weak, and often excluded by symmetry \cite{Fleury68a}. Moreover, our excitation is not magnetic-dipole-active. Even if it were weakly magnetic-dipole-active, we should see this excitations in the geometry with crossed polarizers,\cite{Fleury68a} whereas we see it with parallel polarizers. The magnetic-quadrupole-active excitations in Raman spectra are also considered to be very weak. By this reasoning, we excluded completely the one-magnon scattering as the origin of presence of such a strong electromagnon in our Raman spectra.

There is also a possibility of two-magnon Raman scattering. In general, two-magnon Raman scattering is more likely to occur, because it obeys to looser selection rules. However, two-magnon or two-phonon Raman scattering should weaken on cooling, because these processes are proportional to the population of these quasiparticles which obey to the Bose-Einstein statistics. In our case we see the opposite behavior---the electromagnon intensity noticeably increases on cooling. It is therefore clear that the Raman activity of the electromagnon in ${\rm Ba}{\rm Sr}\rm {CoZnFe}_{11}{\rm Al}{\rm O}_{22}$ must be explained by strong spin-lattice coupling originating from exchange striction.

\subsection{Discussion of the unusually high intensity of the electromagnon}

The intensity of the electromagnon in THz spectra is proportional to the oscillating electric polarization. According to the Eq.~3 and Tab.~I in the main text, the oscillating polarization $\overrightarrow{P}(t)$  is proportional to the ME effect described by the prefactor $\overrightarrow{P}_{i,j}$, inherent to the structure, spin length $S_i$, and spin deviations $\mathrm{\delta}S_j$:

\begin{eqnarray}
 \overrightarrow{P}(t)\propto \overrightarrow{P}_{i,j} \cdot S_i \cdot \mathrm{\delta}S_j.
\end{eqnarray}

In antiferromagnets, electromagnons are usually strong in THz spectra due to the strong ME coupling, but the spin deviations are small due to strong exchange coupling; only extremely intensive pulses can induce larger spin oscillations. However, in our case of a ferrimagnet with frustrated magnetic structures, very different magnetic structures can have very similar energies, therefore spin oscillations can be much stronger than in unfrustrated antiferromagnets. This is why we believe that in our case, $\overrightarrow{P}_{i,j}$ is rather small and $\mathrm{\delta}S_j$ rather large compared to a similarly strong electromagnon in an unfrustrated antiferromagnet observed in THz spectra.

In the case of Raman spectra, the intensity of the electromagnon is probably caused by fluctuations of the oscillating polarization, $\delta\overrightarrow{P}(t)$ from Eq.~(1). As the prefactor $\overrightarrow{P}_{i,j}$ in Eq. (1), given by the ME coupling, is a microscopic parameter inherent to the structure, it does not fluctuate in the first approximation. The high intensity of our electromagnon in the Raman spectra then stems from unusually large spin fluctuations---more precisely, from the fluctuations of $S_i \cdot \mathrm{\delta}S_j$-type terms. This explains why in cases of unfrustrated antiferromagnets, intensities of electromagnons in Raman spectra were low compared to their intensities in THz (or IR) spectra. The case of frustrated ME ferrimagnets is quite rare, which explains why our observation of a strong electromagnon in Raman spectra was the first of such a kind.

We propose that our hypothesis can be tested experimentally in the following way: If the spin vibrations are much larger in our case compared to an unfrustrated antiferromagnet, we should see nonlinear changes in the electromagnon strength depending on the intensity of THz radiation. The required THz electric field should be of the order of 10--100 kV/m; such values are achievable nowadays.

Going deeper into details, we can try to guess why the electromagnon appears weaker in the $a^2$- and $b^2$-polarized Raman spectra than in the $c^2$ spectra. As the oscillating polarization is allowed only in the \textit{c}-direction, electric polarizability in the \textit{ab}-plane is expected to be rather small.

It is important to note that our explanation is just a guess and it needs a theoretical verification. The following theories should be developed: First, our factor-group analysis considered only nonmagnetic space groups, and a factor-group analysis considering magnetic space groups should be worked out. Next, we are reasoning in terms of the electric and magnetic polarizability tensor. However, it is possible that a new response function, that of magnetoelectric polarizability, should be introduced. This would represent an analogy to the magnetoelectric susceptibility, introduced to explain the directional dichroism, which cannot be explained just by using dielectric permittivity and magnetic permeability tensors. Finally, microscopic theories of Raman scattering on electromagnons should be developed for every specific material, using for example a time-dependent density functional theory.
\\
\\
\section{Determination of errors in time-domain THz transmission spectroscopy}

The spectra of the complex refractive index were calculated numerically from those of complex transmittance, which were obtained as ratios of time-windowed sample and reference waveforms, respectively. The time windows were chosen so as to include only the directly passing pulses, omitting the Fabry-P\'{e}rot-like reflections due to the surfaces of the sample.

The numerical method is based on
solving Eq.~(2) in \cite{Kuzel00}.
However, its solutions are unambiguous only for well transparent samples; for samples
exhibiting narrow absorption bands, only the index of absorption
$\kappa(\omega)$ can be determined unambiguously. The refractive index
$n(\omega)$ can be calculated without ambiguities, taking into account the time
delay of the transmitted pulses, only in an interval spanning from the lowest
frequencies up to the first strong and sharp absorption band, whose strong
absorption causes an uncertainty in the phase shift by multiples of $2\pi$.
Thus, one has to choose among several numerical solutions satisfying
Eq.~(2) in \cite{Kuzel00}, separated by a phase shift of $2\pi$, also called branches.
Above the absorption bands, the physically correct ones can be chosen based on the knowledge of the typical shape of a resonance (Lorentz oscillator)---a peak in $\kappa(\omega)$ must be accompanied by an appropriate drop in $n(\omega)$ as the frequency increases. Moreover, the correctness of the spectra can be verified by using the Kramers-Kronig relations---within the limits given by the experimental errors, the real part should transform into the imaginary one and vice versa. In this way, the correct spectra can be found in the whole spectral range.

One has to note also that the overall errors in determining the values of the complex refractive index are not constant in the useful spectral range. The errors are higher near the ends of the useful spectral interval, where the signal intensity is inherently low, and also near the resonances, where the transmitted signal is decreased by the stronger absorption.

In our case, the signal of the source is the highest at $\sim 1\,\rm THz$,
and then decreases to the noise level at $\sim 2.5\,\rm THz$. Therefore, even
the relatively high electromagnon absorption at $\sim 1.2\,\rm THz$ can be quite
well resolved. In contrast, a markedly lower absorption near 2\,THz is near the
limit of the instrument possibilities,
since the sample absorption decreases the transmitted intensity close to the noise level.
\end{document}